\documentclass[12pt]{article}
\usepackage[slantedGreek]{mathptmx}
\usepackage{fullpage}
\usepackage{graphicx}        
\usepackage{amssymb, amsmath,bm}
\usepackage{stmaryrd}
\usepackage{enumitem}
\usepackage{comment}
\usepackage{caption, rotating}
\usepackage[small]{subfigure}
\usepackage[small]{subfigure}
\usepackage{amsfonts}
\usepackage{pdflscape}
\usepackage{xcolor}
\renewcommand{\d}{\text{d}}

\numberwithin{equation}{section}
\usepackage[numbers,square,comma,sort&compress]{natbib}
\graphicspath{{./figures/}} 

\title{Analysis of adiabatic shear coupled to ductile fracture and melting in viscoplastic metals} 

\author{J.D. Clayton$^{1}$\footnote{Email: john.d.clayton1.civ@army.mil} \\ \\
$^1$Terminal Effects Division, Army Research Directorate
\\
DEVCOM ARL, Aberdeen, MD 21005-5066, USA 
}

\date{}

\begin{document}
\graphicspath{{figures/}}
\maketitle

\begin{abstract}
Material failure by adiabatic shear is analyzed in viscoplastic metals that can demonstrate  
up to three distinct softening mechanisms: thermal softening, ductile fracture, and melting.
An analytical framework is constructed for studying simple shear deformation with superposed
static pressure. A continuum power-law viscoplastic formulation is coupled to a ductile
damage model and a solid-liquid phase transition model in a thermodynamically consistent
manner. 
Criteria for localization to a band of infinite shear strain are discussed. An analytical-numerical
method for determining the critical average shear strain for localization and commensurate stress decay is devised.
Averaged results for a high-strength steel agree reasonably well with experimental dynamic torsion data.
Calculations probe possible effects of ductile fracture and melting on shear banding, and vice-versa, including 
influences of cohesive energy, equilibrium melting temperature, and initial defects.
A threshold energy density for localization onset is positively correlated to critical strain and inversely correlated
to initial defect severity. 
Tensile pressure accelerates damage softening and increases defect sensitivity, promoting shear failure.
In the present steel, melting is precluded by ductile fracture for loading conditions
and material properties within realistic protocols.
If heat conduction, fracture, and damage softening are artificially suppressed, melting is confined to a narrow region in the core of the band.
 \end{abstract}

\noindent


\section{Introduction \label{sec1}}

Shear localization is a prevalent failure mode in solid materials that undergo strain-softening mechanisms.
In crystalline metals deformed at high rates, near-adiabatic conditions are obtained, promoting a build up
of local internal energy and temperature from plastic work, in turn leading to thermal softening as
dislocation mobility increases with temperature. A mechanical instability ensues, beyond which
localization into a thin band of very high shear strain and strain rate occurs \cite{wright2002,yan2021}.  Shear banding is often
accompanied by damage mechanisms such as cracking and void growth \cite{cho1990,minnaar1998,fellows2001,fellows2001b,singh2021}, and melting is theoretically
possible if temperatures become large enough \cite{claytonJMPS2024}.
In this work, ``damage''  and ``ductile fracture'' are used to refer changes in local material structure---distinct from phase transformation and deformation twinning and not captured by thermal softening alone in the context of continuum plasticity theory---that induce degradation of local strength.

Experiments and models on adiabatic shear spanning the previous four decades are reviewed in the monograph \cite{wright2002} and more recent article \cite{yan2021}.
Most previous continuum modeling of adiabatic shear localization in metals, be it analytical \cite{staker1981,bai1982,molinari1986,molinari1987,fressengeas1987,molinari1988,wright1990,grady1991,gur2018} or numerical \cite{shawki1983,wright1985,wright1987,cherukuri1995,schoenfeld2003,batra2004,fermencoker2004}, did not explicitly address fracture or melting mechanisms.
However, several continuum damage mechanics-type models for adiabatic shear \cite{zhou1998,longere2003,dolinski2015,dolinski2015b} represented degradation distinctly from purely thermal softening.
More recently, phase-field models \cite{mcauliffe2015,mcauliffe2016,arriaga2017,xu2020,wang2020,samaniego2021,zeng2022}, extending ideas for
depicting brittle fracture and structural transformations in elastic solids \cite{levitas1996,karma2001,levitas2009,claytonPHYSD2011,levitas2011,claytonIJF2014}, have addressed ductile fracture and adiabatic shear localization in
elastic-plastic solids. 
A common element of many models is a contribution of some fraction of energy from plastic work to the net driving force for fracture \cite{mcauliffe2015,mcauliffe2016,arriaga2017,choo2018,wang2020,samaniego2021,zeng2022}, supported
by experiments that show importance of microstructure- and damage-softening relative to thermal softening \cite{rittel2006b,rittel2008}.
Furthermore, many continuum damage and ductile-fracture phase-field models include a threshold inelastic energy density at which damage nucleates \cite{dolinski2010,dolinski2015,dolinski2015b,wang2020,samaniego2021}, as opposed to classical phase-field fracture mechanics wherein
any elastic energy can initiate (diffuse) cracking \cite{claytonIJF2014,claytonPM2015}. 
Timing of cracking or void nucleation relative to localized flow varies among metals with different compositions and microstructures \cite{mebar1983,minnaar1998,fellows2001,fellows2001b,xu2008,singh2021}.
Those cited experiments usually suggest that damage mechanisms accompany or follow localization, rather than precede it, since cracks and voids are scarcely seen outside shear bands in those materials tested.

Direct evidence of melting has been reported within dynamic shear bands in various titanium and steel alloys \cite{mebar1983,wang2009,li2011,healy2015}.
Known prior analyses of adiabatic shear did not include the thermodynamics of melting, even though local temperatures far exceeding
the melting point of iron or steel can be predicted inside the shear band \cite{claytonJMPS2024,molinari1986,molinari1988,bronkhorst2006}.
Phase-field models combining mechanics of elasticity and thermodynamics and kinetics of melting and solidification exist \cite{levitas2011b,hwang2014,hwang2016,goki2024}, but these do not consider plastic deformation or adiabatic shear.
A continuum model described flowing material within a shear band using a Bingham model with Newtonian viscosity \cite{saga2018}, at temperatures below the ambient melt temperature. 
Therein, the calibrated viscosity was so low for three different metallic systems that the constant, rate-independent part of the shear stress dominated.

A recent theoretical study \cite{claytonJMPS2024} advanced the analyses of Molinari and Clifton \cite{molinari1986,molinari1987,molinari1988}
to account for influences of superposed pressure, external magnetic fields, and solid-solid phase transformations on adiabatic shear localization.
Results showed how loading conditions and solid-solid phase transformations can promote or inhibit strain localization in iron and a high-strength Ni-Cr steel.
Herein, treatments of Refs.~\cite{claytonJMPS2024,molinari1986,molinari1987,molinari1988} are further extended to account for damage 
(i.e., ductile shear fracture) and melting (i.e., solid-liquid transformation) processes. 
Criteria for localization into a band of infinite shear \cite{molinari1986} are analyzed, and methods for calculating 
the average shear strain at which localization ensues are developed.
The latter require numerical iteration and numerical integration, as closed-form expressions for critical strain cannot be derived analytically.
The ductile fracture component of the model further addresses the additional ``average'' shear strain accommodated by the sample
after localization, accounting for the effective shear displacement jump across the band whose shear strain approaches infinity and width approaches zero. This allows for representation of a gradual softening behavior in the macroscopic shear stress versus shear strain
response as regularly witnessed experimentally \cite{dolinski2010,dolinski2015b,fellows2001,fellows2001b,marchand1988}.  Overly abrupt drops to zero stress upon localization \cite{molinari1986,molinari1987,claytonJMPS2024} were criticized
by Molinari and Clifton \cite{molinari1987} as a notable deficiency of their original approach, therein speculated a result of poor resolution of defects.

As assumed or justified in prior works \cite{molinari1987,wright1994,wright2002,claytonJMPS2024} for materials (e.g., steels) and average strain
rates (e.g., $10^3-10^4$/s) of present interest,
regularization mechanisms of heat conduction and inertia are omitted to enable a tractable analysis without recourse to advanced
numerical methods such as finite elements or finite differences.
Similarly, to allow localization into a damaged region of infinitesimal thickness, the ductile fracture component of the formulation
omits gradient regularization of phase-field theory \cite{claytonPRE2024}, and Newtonian viscosity of molten material is likewise omitted in the limit of singular surfaces \cite{morro1980,molinari1986,molinari1987,claytonJMPS2024}.
Any remaining (implicit) regularization is furnished by strain-rate sensitivity \cite{needleman1988}.
Results are thus interpreted as a limiting case: predictions would be expected to underestimate conditions for localization
given the absence of other regularization mechanisms \cite{wright1994,wright2002}. 
An initial defect (e.g., strength perturbation) of greater intensity than
 imposed or predicted here and in Refs.~\cite{molinari1986,molinari1987,claytonJMPS2024} would be needed to instill the same post-peak shear strain.

This article consists of six more sections. In \S2, a general 3-D continuum framework is outlined, including constitutive fundamentals and thermodynamics.
In \S3, specialization of the framework to simple shear and pressure loading is undertaken. Constitutive model components for
viscoelasticity, ductile fracture, and melting are introduced in this context.
In \S4, localization criteria are examined, and methods of calculation of critical shear strain and average stress-strain response are explained. 
In \S5, properties and results are reported for a high-strength steel and compared to experimental observation.
In \S6, effects of variations of material parameters on localization behaviors are explored.
In \S7, conclusions consolidate the main developments.
Standard notation of continuum mechanics is used (e.g., Refs.~\cite{wright2002,claytonNCM2011}), with vectors and tensors in bold
font and scalars and scalar components in italics. A single Cartesian frame of reference is sufficient for this work.

\section{3-D formulation \label{sec2}}

The general constitutive framework combines elements from Refs.~\cite{claytonCMT2022,claytonZAMM2024,claytonJMPS2024,claytonJMPS2021,mcauliffe2015,levitas2011,levitas2011b,hwang2014,hwang2016}.
Electromagnetic effects considered in Refs.~\cite{claytonCMT2022,claytonZAMM2024,claytonJMPS2024} are excluded,
but phase-field concepts for fracture \cite{claytonJMPS2021,mcauliffe2015,levitas2011,claytonJMPS2021,choo2018} and melting \cite{levitas2011b,hwang2014,hwang2016} are now added.
The material is isotropic in both solid polycrystalline and liquid amorphous states, and is assumed fully solid in its initial configuration.

Inertial dynamics, heat conduction, and surface energies are 
included the complete 3-D theory, as are thermal expansion and finite elastic shear strain.
These features are retained in \S2 for generality and to facilitate identification and evaluation of successive approximations made later.
Furthermore, retainment of such physics in the general formulation will allow a consistent implementation of the complete nonlinear
theory in subsequent numerical simulations, for potential future comparison to the results of semi-analytical calculations reported in 
\S5 and \S6.

\subsection{Constitutive model}
Denote the motion of a material particle at reference position $\bf{X}$ by spatial coordinates ${\bf x} $ = $ {\bm \varphi } ({\bf X},t)$.
Gradients with respect to $\bf{X}$ and $\bf{x}$ are written $\nabla_0(\cdot)$ and $\nabla(\cdot)$, and a superposed dot denotes the time
derivative at fixed $\bf{X}$. The deformation gradient $\bf{F}$ and its determinant $J$ are decomposed into a product of
three terms \cite{claytonNCM2011,levitas1998,claytonIJP2003}, with ${\bf F}^E$ thermoelastic deformation, ${\bf F}^P$ plastic deformation
from dislocation motion, and ${\bf F}^I$ the total inelastic deformation from damage and melting:
\begin{align}
\label{eq:defgrad} 
& {\bf F} = \nabla_0 {\bm \varphi} = {\bf F}^E {\bf F}^I {\bf F}^P, 
\qquad 
J = \det {\bf F} = \det {\bf F}^E \det {\bf F}^I \det{\bf F}^P = J^E J^I J^P > 0.
\end{align}
None of ${\bf F}^E, {\bf F}^I, {\bf F}^P$ need be individually integrable to a vector field \cite{claytonDGKC2014}. Denote dimensionless order parameter fields for melting by $\xi({\bf X},t) \in [0,1]$ and damage by $\phi({\bf X},t) \in [0,1]$, where
\begin{align}
\label{eq:ops}
\xi  \begin{cases}
& =  0 \quad \leftrightarrow \quad \text{solid material}, \\
& \in (0,1) \, \leftrightarrow \, \text{mixed phase region},  \\
& = 1 \quad \leftrightarrow \quad \text{liquid material};
\end{cases}
\quad
\phi  \begin{cases}
& =  0 \quad \leftrightarrow \quad \text{undamaged material}, \\
& \in (0,1) \, \leftrightarrow \, \text{partially degraded material},  \\
& = 1 \quad \leftrightarrow \quad \text{fully degraded material}.
\end{cases}
\end{align}
Deformations of solid and liquid phases are not tracked individually at a point $\bf X$, so each is effectively assigned the same
overall deformation gradient $\bf F(\bf X,t)$ at time $t$.
Accordingly, $\xi$ is interpreted equivalently as the local mass fraction or local volume fraction of molten material.
On the other hand, $\phi$ need not represent the local volume fraction of voids or free volume, but rather is a generic
indicator of local strength loss.

In a body of reference volume $\Omega_0$, reference conditions are $\xi({\bf X}) = \phi({\bf X}) = 0 \, \forall \, {\bf X} \in \Omega_0$. Calculations do not consider bodies with initial liquid phases or with initial damage represented by nonzero values of
$\xi({\bf X},0)$ and $\phi({\bf X},0)$, respectively. If the material contains initial defects associated with damage entities (e.g., pores, micro-cracks, or other intrinsically weak zones),
$\phi({\bf X},t)$ represents only the additional damage incurred after loading begins, potentially nonzero only for $t > 0^+$.  In \S3.3, a relationship between initial strength perturbation $\delta \chi_0 (\bf{X})$ and an initial damage variable physically analogous to, but mathematically distinct from, 
$\phi({\bf X},0) = 0$, is introduced.

Deformation ${\bf F}^I$ is a function of state (i.e., of order parameters), plasticity is isochoric, and symmetric thermoelastic deformation is measured by ${\bf C}^E$
with isochoric part $\tilde{\bf C}^E$:
\begin{align}
\label{eq:FxiJp}
{\bf F}^I = {\bf F}^I(\xi,\phi); \qquad J^P = \det {\bf F}^P = 1; \qquad {\bf C}^E = ({\bf F}^E)^{\rm T}{\bf F}^E, \quad \tilde{\bf{C}}^E = J^{E \, -2/3} {\bf C}^E. 
\end{align}

Denote by $\chi({\bf X}, t)$ a scalar internal state variable field associated with plastic deformation processes (e.g., a dimensionless measure of dislocation density).
Let $\theta({\bf X}, t)$ be absolute temperature.
Helmholtz free energy per unit reference volume of the solid is $\psi$, consisting of thermoelastic strain energy $W$, thermal energy $Q$, microstructure energy $R$,
 and gradient surface energy $\Lambda$:
\begin{align}
\label{eq:psi}
\psi({\bf C}^E, \theta, \xi, \phi, \chi, \nabla_0 \xi, \nabla_0 \phi) = 
W({\bf C}^E, \theta, \xi, \phi) + Q( \theta, \xi) + R(\xi, \phi, \chi) + \Lambda(\xi,\phi, \nabla_0 \xi, \nabla_0 \phi).
\end{align}

Denote by $\theta_0$ a reference temperature and $\Delta \theta = \theta - \theta_0$. Non-degraded isothermal bulk modulus $B_0$ is assumed the same in solid and liquid phases for neutral and compressive states wherein $\ln J^E \leq 0$. Volumetric coefficient of thermal expansion $A_0$ is assumed the same in solid and liquid.
The non-degraded shear modulus of the solid is $\mu_0$. Thermoelastic strain energy combines a logarithmic equation of state \cite{claytonIJES2014,claytonNEIM2019} with a polyconvex shear contribution \cite{claytonPRE2024}:
\begin{align}
\label{eq:W}
& W(J^E,\tilde{\bf C}^E,\theta,\xi,\phi) = {\textstyle{\frac{1}{2}}} B(\xi,\phi) (\ln J^E)^2 [ 1 -   {\textstyle{\frac{1}{3}}} (B'_0 - 2) \ln J^E ] \nonumber \\ & \qquad \qquad \qquad \qquad \qquad \qquad \qquad \qquad
- A_0 B \Delta \theta  \ln J^E 
+  {\textstyle{\frac{1}{2}}} \mu(\xi,\phi) ( {\rm tr} \, \tilde{\bf C}^E - 3).
\end{align}

Let $\iota^\xi(\xi) \in [0,\iota^\xi_1]$ be an interpolation function between solid and liquid states satisfying $\iota^\xi(0) = 1, \iota^\xi(1) = \iota^\xi_1 \in [0,1]$.
Let $\bar{\omega}(\phi,J^E) = 1 +  [{\omega}(\phi) - 1] \mathsf{H}(\ln J^E) \in [0,1]$ be a degradation function satisfying ${\omega}(0) = 1, {\omega}(1) = \omega_1 \in [0,1]$, where $\mathsf{H}(\cdot)$ is the left-continuous Heaviside function. 
Moduli are interpolated as follows,
noting as $\xi \rightarrow 1$ and $\phi \rightarrow 1$, $B \rightarrow B_0 \omega_1 \iota^\xi_1 $ in tension and $\mu \rightarrow \mu_0 \omega_1 \iota^\xi_1$:
\begin{equation}
\label{eq:moduli}
B(\xi,\phi,J^E) = B_0 \bar{\omega}(\phi,J^E) [ 1 +\{ \iota^\xi (\xi) - 1 \} \mathsf{H}(\ln J^E) ], \qquad \mu(\xi,\phi) = \mu_0 {\omega}(\phi) \iota^\xi(\xi).
\end{equation}
Constants $\omega_1$ or $\iota^\xi_1$, when nonzero, enable remnant strength when 
a material element is fully fractured or melted, and
$B = B_0$ in compression. 
Cauchy pressure $p$ and deviatoric Cauchy stress $\tilde{ \bm \sigma}$ then follow from \eqref{eq:W}, with $\tilde{\mathbf B}^E$
the spatial deviatoric lattice strain:
\begin{align}
\label{eq:lattp}
& \bm{\sigma} = \tilde{\bm \sigma} - p {\bf 1}; \quad p = - \frac{1}{3}{\rm tr} {\bm \sigma} = - \frac{1}{J} \frac{\partial \psi}{\partial \ln J^E} = 
- \frac{B}{J} \{   \ln J^E  [ 1 - {\textstyle{\frac{1}{2}}}  (B'_0 - 2) \ln J^E] - A_0  \Delta \theta \};
\\
\label{eq:lattsig}
& \tilde{ \bm \sigma}= \frac{2}{J} {\bf F}^E \frac{ \partial  \psi }{\partial  \tilde{\bf C}^E}: \frac{\partial  \tilde{\bf C}^E}{\partial {\bf C}^E} ({\bf F}^E)^{\rm T} = \frac{\mu}{J} \tilde{\bf B}^E, \quad
\tilde{\bf B}^E = (J^E)^{-2/3} {\bf F}^E ({\bf F}^E)^{\rm T} - \frac{1}{3} {\rm tr} \, [ (J^E)^{-2/3} {\bf F}^E ({\bf F}^E)^{\rm T} ] {\bf 1}.
\end{align}

Specific heat per unit volume $c_V$ is assumed the same for solid and liquid phases and is idealized as independent of $\theta$.
Latent heat of fusion per unit reference volume of the solid is the constant $h_T$, positive for the usual case of higher energy of
the liquid than the solid coexisting at the same $(p,\theta)$. Herein $h_T \geq 0$. An equilibrium transformation temperature is the constant $\theta_T$. At this temperature, in the absence of elastic deformation, stored energy of microstructure, and gradient surface energy, the free energy densities of solid and liquid phases are equal, whereby $\theta_T$ is interpreted as the melt temperature.
Thermal energy is 
\begin{align}
\label{eq:Q}
Q (\theta,\xi) = - c_V [ \theta \ln (\theta / \theta_0) - \Delta \theta] +  [ (h_T / \theta_T) (\theta - \theta_T)] \iota^\xi(\xi).
\end{align}
Denote by $\theta_I \leq \theta_T$ a melt instability temperature and the constant $A_\xi = 3 h_T (1 - \theta_I / \theta_T) \geq 0$ \cite{levitas2011b} .

Denote by $E_C \geq 0$ the cohesive energy per unit reference volume \cite{levitas2011,claytonPRE2024}, presumed matching in solid and liquid phases here for simplicity. 
More generally, $E_C$ could depend on $\xi$ and $\theta$, to allow for different, temperature-dependent cavitation thresholds 
in solid and liquid. However, data to justify such generalizations, incurring additional driving and resistive forces
for melting and fracture and increasing model complexity and computational burden, appear difficult to measure and do not seem to exist for the material analyzed in \S5.
Microstructure energy is
\begin{align}
\label{eq:Ren}
R(\xi,\phi,\chi) = \mu_0 [1-r_0^\xi \{1-\iota^\xi(\xi)\}] \bar{R}(\chi) + A_\xi \xi^2( 1- \xi)^2 + E_C f_\phi (\phi),
\end{align}
with $\bar{R}$ a dimensionless function of
$\chi$. 
The first term on the right in \eqref{eq:Ren}, for stored energy of dislocations, is scaled by the shear modulus \cite{claytonNCM2011}, the second is a double-well 
for phase boundaries \cite{levitas2011b,hwang2014,hwang2016}, and the third, with $f_\phi (\phi) \in [0,1]$ a dimensionless function, is for homogeneous fracture \cite{claytonJMPS2021,levitas2011}.
Constant $r_\xi^0 \in [0,1]$ is the fraction of stored energy of cold work released upon melting.

Let $\Gamma_\xi$ and  $\Gamma_\phi$ be surface energies for solid-liquid boundaries and cracks, respectively, and let $l_\xi, l_\phi$ be regularization length constants.
Gradient energy of internal surfaces is then the usual sum of quadratic forms \cite{claytonJMPS2021,levitas2011b}, where here
surface energies can generally depend on order parameters:
\begin{align}
\label{eq:graden}
\Lambda(\xi, \phi, \nabla_0 \xi, \nabla_0 \phi) = \Gamma_\xi(\phi) l_\xi |\nabla_0 \xi|^2 + \Gamma_\phi(\xi) l_\phi |\nabla_0 \phi |^2.
\end{align}
Surface tension \cite{levitas2011b,hwang2014,hwang2016} is omitted from $R$ and $\Lambda$ for brevity and simplicity; it could be included by multiplying right sides of \eqref{eq:Ren} and \eqref{eq:graden} by $J$ and replacing $\nabla_0(\cdot)$ with $\nabla(\cdot)$ in \eqref{eq:graden}.

For later use, define first Piola-Kirchhoff stress ${\bf P}$, elastic second Piola-Kirchhoff stress ${\bf S}$, Mandel stress $\bar{\bf S}$, and the thermoelastic entropy constitutive relation as 
\begin{align}
\label{eq:PK1}
& {\bf P} = J {\bm \sigma} {\bf F}^{- \rm{T}}, \quad
{\bf S} = J ({\bf F}^E)^{-1} {\bm \sigma} ({\bf F}^E)^{- \rm{T}} = 2 \, {\partial \psi}/{\partial {\bf C}^E}, \quad
\bar{ \bf S} = {\bf C}^E {\bf S}, 
\quad
 \eta = -\partial \psi / \partial \theta.
\end{align}
With $\eta$ and $U$ being entropy and internal energy per unit volume, ${\bf q}$ the referential heat flux, and $\kappa \geq 0$ isotropic Fourier conductivity generally dependent on temperature, phase, and damage:
\begin{equation}
\label{eq:thermoids}
U = \psi + \theta \eta, \qquad {\bf q}(\xi,\phi,\theta, \nabla_0 \theta)  = - \kappa (\xi,\phi,\theta) \nabla_0 \theta.
\end{equation}

\subsection{Balance laws and dissipation}
Standard local forms \cite{claytonNCM2011,claytonNEIM2019} for conservation of mass and momentum are invoked,
where $\rho_0$ and $\rho$ are referential (solid, undeformed and undamaged) and spatial (deformed, possibly molten, and possibly damaged) mass densities, 
${\bf b}$ is body force per unit mass, and $\bm{\upsilon} = \dot {\bf x}$ is particle velocity:
\begin{align}
\label{eq:balances}
\rho_0 = \rho J, \qquad \nabla \cdot {\bm \sigma} + \rho {\bf b} = \rho \dot{\bm \upsilon}, \qquad {\bm \sigma} = {\bm \sigma}^{\rm T}.
\end{align}
Let ${\bf n}_0$ be the unit outward normal to material body $\Omega_0$ on external boundary $\partial \Omega_0$. Global forms of the balance of energy and entropy inequality are as follows, the first extending classical continuum mechanics to account for surface energetics of order parameters \cite{gurtin1996,claytonJMPS2021,claytonPRE2024}:
\begin{align}
\label{eq:1stglob}
& \frac{\d }{\d t} \int_{\Omega_0} U \d \Omega_0 + \frac{\d }{\d t} \int_{\Omega_0} \frac{\rho_0}{2} | {\bm \upsilon}|^2 \d \Omega_0 = 
\oint_{\partial \Omega_0} ({\bf P} \cdot {\bf n}_0) \cdot {\bm \upsilon} \, \d \partial \Omega_0
- \oint_{\partial \Omega_0} ({\bf q} \cdot {\bf n}_0) \, \d \partial \Omega_0  \nonumber
\\ & \qquad \qquad \qquad \qquad \qquad
+ \oint_{\partial \Omega_0} \left( \frac{\partial \psi}{\partial \nabla_0 \xi} \cdot {\bf n}_0 \right) \dot{\xi} \, \d \partial \Omega_0
+ \oint_{\partial \Omega_0} \left( \frac{\partial \psi}{\partial \nabla_0 \phi} \cdot {\bf n}_0 \right) \dot{\phi} \, \d \partial \Omega_0, \\
\label{eq:2ndglob}
& \frac{\d }{\d t} \int_{\Omega_0} \eta d \Omega_0
+ \oint_{\partial \Omega_0} \frac{{\bf q} \cdot {\bf n}_0}{\theta} \, \d \partial \Omega_0 \geq 0.
\end{align}
The divergence theorem, \eqref{eq:balances}, and differentiability gives local versions of \eqref{eq:1stglob} and \eqref{eq:2ndglob}:
\begin{align}
\label{eq:1stloc}
& \dot{U} = {\bf P}:\dot{\bf F} - \nabla_0 \cdot {\bf q} + \nabla_0 \cdot 
[ (\partial \psi / \partial \nabla_0 \xi) \dot{\xi} + ( \partial \psi / \partial \nabla_0 \phi) \dot{\phi} ], \\
\label{eq:2ndloc}
& \theta \dot{\eta} + \nabla_0 \cdot {\bf q} - ( {\bf q} \cdot \nabla_0 \theta)/\theta \geq 0.
\end{align}
Define the internal dissipation as follows, applying the first of \eqref{eq:thermoids} and \eqref{eq:1stloc}:
\begin{align}
\label{eq:Dint}
\mathfrak{D} = \theta \dot{\eta} + \nabla_0 \cdot {\bf q} = {\bf P}:\dot{\bf F} + \nabla_0 \cdot 
[  (\partial \psi / \partial \nabla_0 \xi) \dot{\xi} + (\partial \psi / \partial \nabla_0 \phi) \dot{\phi}  ]
-\dot{\psi} - \dot{\theta} \eta.
\end{align}
The second law \eqref{eq:2ndloc} is, from \eqref{eq:defgrad}, \eqref{eq:FxiJp}, \eqref{eq:PK1}, \eqref{eq:thermoids}, and chain-rule differentiation of \eqref{eq:psi},
\begin{align}
\label{eq:2nd2}
\mathfrak{D} & + \frac{\kappa}{\theta} |\nabla_0 \theta|^2 \geq 0; \qquad
\mathfrak{D} =  [( {\bf F}^I)^{\rm T} \bar{\bf S} ({\bf F}^I)^{\rm -T}]:{\bf L}^P + \varsigma \dot{\xi}  + \zeta \dot{\phi} + \vartheta \dot{\chi},
\qquad {\bf L}^P = \dot{\bf F}^P {\bf F}^{P -1}; 
\\ 
\label{eq:varsigma}
 \varsigma & = - \partial \psi / \partial \xi + \nabla_0 \cdot (\partial \psi / \partial \nabla_0 \xi) + {\bf P}:(\partial {\bf F} / \partial \xi) \nonumber \\
& = - \partial (W + Q + R + \Lambda) / \partial \xi + 2 l_\xi \nabla_0 \cdot (\Gamma_\xi  \nabla_0 \xi  )
+  [\bar{\bf S} ( {\bf F}^I)^{- {\rm T}}]:(\partial {\bf F}^I / \partial \xi),
\\
\label{eq:zeta}
 \zeta & = - \partial \psi / \partial \phi + \nabla_0 \cdot (\partial \psi / \partial \nabla_0 \phi) + {\bf P}:(\partial {\bf F} / \partial \phi) \nonumber \\
& = - \partial (W + R + \Lambda) / \partial \phi + 2 l_\phi \nabla_0 \cdot (\Gamma_\phi \nabla_0 \phi )
+ [\bar{\bf S} ( {\bf F}^I)^{- {\rm T}}]:(\partial {\bf F}^I / \partial \phi),
\\ 
\label{eq:nu}
 \vartheta & = - \partial \psi / \partial \chi = - \mu_0 [1-(1-\iota^\xi)r_0^\xi  ] \partial {\bar{R}} / \partial \chi.
\end{align}
Expansion of $\dot{\eta} = - \frac{\partial \dot{\psi}}{ \partial \theta}$ with \eqref{eq:Q}, and \eqref{eq:1stloc}, then give the temperature rate, noting $c_V = - \theta \frac{\partial^2 \psi }{\partial \theta^2}$:
\begin{align}
\label{eq:Tdot}
c_V \dot{\theta} = \mathfrak{D} + h_T \frac{\theta}{ \theta_T} \frac{ \d \iota^\xi }{ \d \xi} \dot{\xi} - 
A_0 \theta \biggr{[} B  \frac{\dot{J}^E}{ J^E } + \ln J^E \biggr{\{} \frac{\partial B}{ \partial \xi} \dot{\xi} + \frac{\partial B}{ \partial \phi} \dot{\phi} \biggr{\}}
\biggr{]}
+ \nabla_0 \cdot (\kappa \nabla_0 \theta).
\end{align}

\section{Dynamic shear with pressure \label{sec3}}

The problem analyzed in \S3--\S6 is similar to that of Ref.~\cite{claytonJMPS2024}.
However, magnetization and magnetic fields considered in that work are omitted here,
as are solid-solid phase transformations.
Instead, solid-liquid transformations and ductile fracture are now included.
It is possible to posit a reduced set of governing equations for shear band analysis from the outset,
without referencing the full 3-D constitutive model of \S2 that encompasses thermoelasticity, conduction, 
gradient regularization, and
inertia. However, that full formulation is used as a starting point here to clarify the simplifying assumptions needed for mathematically tractable limit
analysis in later sections.

\subsection{Geometry, loading conditions, and governing equations}

\begin{figure}
\centering
\includegraphics[width=0.9\textwidth]{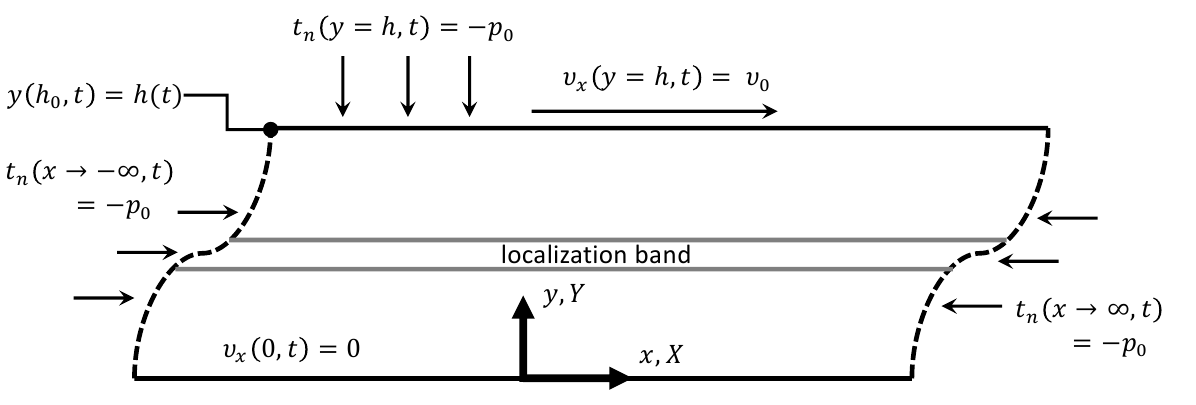}
\caption{Boundary value problem for simple shear with superposed static pressure: tangential velocity at $y=h$ is $\upsilon_0$, normal traction $t_n$ is $p_0$, including $t_n (z \rightarrow \pm \infty,t) = -p_0$. Slab is infinite in $X,x,Z,z$ directions; initial (current) height of slab is $h_0$ ($h$). 
Slab is thermally insulated: $\partial \theta / \partial y = 0$ at $y = 0$ and at $y = h$ for $t \geq 0^+$.
Initial variable $\chi_0(y) = \chi(y,t = 0)$ and temperature $\theta(y,t=0)$ can vary modestly over $y \in (0,h)$.
\label{fig1}}
\end{figure}

Shown in Fig.~\ref{fig1} is the transient boundary value problem of present study.
Let $(X,Y,Z)$ and $(x,y,z)$ be Cartesian reference and spatial coordinates.
The material domain $\Omega_0$ is of initial height $h_0$ and current height $h(t)$.
The slab is infinitely extended in $X,x$- and $Z,z$- directions. 
The following mixed boundary conditions are invoked. Denote by $\upsilon_x = \upsilon_0$ a constant velocity at $y = h$, with vanishing velocity $\upsilon_x = 0$ at $y=0$. Constant pressure field $t_n = - p_0$ is applied as normal traction on all boundaries for $t \geq 0^+$.  
Though not depicted in Fig.~\ref{fig1}, tangential traction from $\sigma_{xy}$ at each cross section $x = \text{constant}$ exists. Letting $(X,Y,Z)$ be coordinates \textit{minus} application of $p_0$, $h_0$ can be unequal to the $Y$ value of the planar boundary of the stress-free slab due to compressibility (i.e., $h_0$ depends on $p_0$).
Pressure $p_0$ is applied slowly so is treated as an isothermal and quasi-static pre-loading. Then for dynamic shearing when $t \geq 0^+$, $y=0$ and $y=h$ are thermally insulated.
 
For $t > 0^+$, while the body is shearing, $h$ may displace from $h_0$ to maintain $p_0 = \text{constant}$ at $y= h$.  This would occur if a volume-changing structural change such as melting or void growth is driven by shear.
Note $\upsilon_0$ is constant for $t \geq 0^+$: acceleration from a resting state to an initial velocity gradient is not modeled.
For $t \geq 0^+$, with infinite $x,z$ boundaries, spatial fields depend at most only on $(y,t)$, a common assumption in analytical studies of shear bands \cite{wright2002,molinari1986,molinari1987,fressengeas1987,le2018}:
\begin{align}
\label{eq:1D}
{\bm \upsilon} = {\bm \upsilon}(y,t), \qquad {\bf F} = {\bf F}(y,t), \qquad {\bm \sigma } = {\bm \sigma}(y,t), \qquad \theta = \theta(y,t), \qquad [ t \geq 0^+].
\end{align}
Any single localization band, within which ${\bf F}$ differs from most of the rest of the domain, is oriented as in Fig.~\ref{fig1}, but multiple parallel bands are allowable.
Properties and initial temperature are constant with respect to $x$ and $z$ but need not so with respect to $y$. Perturbations in initial strength and temperature \cite{molinari1988} can trigger localization at critical $y$ locations. 

Reference configuration covered by $(X,Y,Z)$ is a model construct for analysis and need not be attained in experimental practice. Pressure $p_0$ is removed instantly at fixed $\theta$ from an unsheared state, before $\upsilon_x = \upsilon_0$ is applied. By construction, ${\bf F} = {\bf F}^E = {\bf F}^I = {\bf F}^P = {\bf 1}$, ${\bm \sigma} = {\bf 0}$,
and $\phi = \xi = 0$ in this reference configuration.  Almost everywhere, $\theta \approx \theta_0 = \text{constant}$ and 
$\chi \approx \chi_0 = \text{constant}$.  
Initial perturbations in $\theta$ and $\chi$ are permitted, with magnitudes so small that initial perturbations in $J$, $p$, $\phi$, and $\xi$ can be omitted.

Since the constitutive model of \S2 is isotropic, the only nonzero Cauchy stress components are $\sigma_{xx}$, $\sigma_{yy}$, $\sigma_{zz}$, and $\sigma_{xy} = \sigma_{yx}$, and the nonzero heat flux component is $q_Y$. 
Taking ${\bf b} = {\bf 0}$ henceforth and using \eqref{eq:1D}, the second of \eqref{eq:balances} and heat conduction expressions in \eqref{eq:thermoids} and  \eqref{eq:Tdot} are
\begin{align}
\label{eq:mombal1D}
&  \frac{\partial \sigma_{xy}}{ \partial y}  = \rho \dot{\upsilon}_x,
\qquad  
\frac{\partial \sigma_{yy}}{ \partial y }= \rho \dot{\upsilon}_y;
\\
\label{eq:heatcon}
&  q_Y  = -  \kappa F_{yY} \frac{\partial \theta} {\partial y},
\qquad
\nabla_0 \cdot \kappa \nabla_0 {\theta} =   F_{yY} \frac{\partial}{\partial y} \left( \kappa F_{yY} \frac {\partial \theta}{\partial y} \right).
\end{align}

Regarding the deformation gradient, non-vanishing components of ${\bf F}(y(Y,t),t)$ are $F_{xX}$, $F_{yY}$, $F_{zZ}$ and $F_{xY}$.
For generic function $f = f(y(Y,t),t)$ (e.g., \eqref{eq:1D}), $\partial f / \partial Y =  (\partial f / \partial y) F_{yY} $,
and $\partial f / \partial X = \partial f / \partial Z = 0$.
From the second of \eqref{eq:1D}, compatibility conditions \cite{claytonDGKC2014} $\nabla_0 \times {\bf F} = {\bf 0}$ necessitate
\begin{align}
\label{eq:compat1D}
\partial F_{xX} / \partial Y = \partial F_{xY} / \partial X = 0, \qquad
\partial F_{zZ} / \partial Y = \partial F_{zY} / \partial Z = 0.
\end{align}

Forms of ${\bm \varphi}$, ${\bf F}^E$, ${\bf F}^\xi$, and ${\bf F}^P$ satisfying \eqref{eq:compat1D} are postulated a priori \cite{claytonJMPS2024}, based on Fig.~\ref{fig1}.
The reference configuration is unstressed, and ${\bf F}^E$ includes the isothermal elastic volume change from initial pressurization by $p_0$. Motion ${\bf x} = {\bm \varphi}({\bf X},t)$ and deformation gradient terms, written in  $3 \times 3$ matrix format, are
\begin{align}
\label{eq:varphi}
& x = x(X,Y,t), \quad y = y(Y,t), \quad z = z(Z,t); 
\qquad \gamma = J^I \gamma^E + \gamma^I + \gamma^P;
\\ 
\label{eq:FE}
& [{\bf F}^E(Y,t)] = (J^{E}(t))^{1/3} 
\begin{bmatrix}
& 1 &  \gamma^E(Y,t) & 0 \\
& 0 & 1 & 0 \\
& 0 & 0 & 1
\end{bmatrix},
\quad
 [{\bf F}^I (Y,t)] = 
\begin{bmatrix}
& 1 &  \gamma^I(Y,t) & 0 \\
& 0 & J^I(Y,t) & 0 \\
& 0 & 0 & 1
\end{bmatrix},
\\
\label{eq:FP}
& [{\bf F}^P (Y,t)] = 
\begin{bmatrix}
& 1 &  \gamma^P(Y,t) & 0 \\
& 0 & 1 & 0 \\
& 0 & 0 & 1
\end{bmatrix},
\quad
 [{\bf F} (Y,t)] = 
\begin{bmatrix}
& 1 &  \gamma(Y,t) & 0 \\
& 0 & J^I(Y,t) & 0 \\
& 0 & 0 & 1
\end{bmatrix}
 (J^{E}(t))^{1/3} .
\end{align}
The thermoelastic term ${\bf F}^E$ contains spatially homogeneous volume change $J^E$ and shear $\gamma^E$.  
 The total shear \textit{after} thermoelastic volume change in the last of \eqref{eq:FP} is $\gamma$. Plastic deformation is a simple shear $\gamma^P$.
Inelastic deformation from damage and melting consists of simple shear $\gamma^I$ and strain normal to the band,
with volume change $J^I$, both interpolated as follows \cite{claytonJMPS2021,levitas2011b,claytonPHYSD2011}:
\begin{align}
\label{eq:gammaI}
& \gamma^I(\xi,\phi) = \gamma^\xi(\xi)+ \gamma^\phi(\phi) ,
 \quad \gamma^\xi = \gamma^\xi_0 (1 - \iota^\xi (\xi)), 
\quad \gamma^\phi = \gamma^\phi_0 (1 - \iota^\phi (\phi));
 \\
 \label{eq:JI}
& J^I(\xi,\phi) = J^\xi(\xi) J^\phi(\phi) =  \frac{J^\xi (\xi) }{1/J^\phi(\phi)}  = \frac{1 + \delta^\xi_0 (1 - \iota^\xi (\xi))}{1 - \delta^\phi_0 (1 - \iota^\phi(\phi))}; 
\qquad (1-\iota_1^\xi) \delta^\xi_0  = \frac{\rho^{(0)}}{ \rho^{(1)}} - 1.
\end{align}
Interpolation functions between solid and liquid, and between undamaged and failed states, are 
\begin{align}
\label{eq:interpolators}
& \iota^\xi (\xi) \in [0,\iota^\xi_1 ], \qquad \iota^\phi (\phi) \in [0,\iota^\phi_1]; 
\nonumber
\\
& \iota^\xi(0) = \iota^\phi(0) = 1; \qquad \iota^\xi(1) = \iota^\xi_1 \in [0,1], \quad \iota^\phi(1) = \iota^\phi_1 \in  [0,1].
\end{align}
Material constants are $\gamma^\xi_0, \gamma^\phi_0, \delta^\xi_0$, and $\delta^\phi_0$, any of which can be zero, positive, or negative. These constants are multiplied by $1 - \iota^\xi_1$ or $1-\iota^\phi_1$ to
give any structure change-induced shear strains and volume changes at $\xi = 1$ or $\phi = 1$. Mass densities of solid and fully liquid states are $\rho^{(0)}$ and $\rho^{(1)}$, respectively.
When damage manifests physically as induced cavities or pores, 
then the quantity $(1-\iota^\phi_1) \delta^\phi_0$ can be interpreted as the void volume fraction
at the locally failed state. Notice that $J^\phi$ accounts only for volume changes
induced during deformation, and not for effects of initial defects (e.g., $J^\phi$ does not account for initial
porosity independent of $\phi$ that is implicitly included in $\rho^{(0)}$).

Inserting ${\bf F}^E$ of \eqref{eq:FE}, the Cauchy stress of \eqref{eq:lattp} and \eqref{eq:lattsig} becomes
\begin{align}
\label{eq:Cauchymatrix}
& [ {\bm \sigma} ] = 
\frac{1}{J^E J^I}
\begin{bmatrix}
& -\tilde{p} + \frac{2}{3} \mu (\gamma^E)^2 & \mu \gamma^E & 0 \\
&  \mu \gamma^E  &  -\tilde{p} - \frac{1}{3} \mu (\gamma^E)^2 & 0 \\
& 0 & 0 & -\tilde{p} - \frac{1}{3} \mu (\gamma^E)^2 
\end{bmatrix},
\\
\label{eq:barp}
& \tilde{p}(J^E,\theta) = - {B} \{   \ln J^E  [ 1 - {\textstyle{\frac{1}{2}}}  (B'_0 - 2) \ln J^E] - A_0  (\theta - \theta_0) \}.
\end{align}

Three approximations are introduced to simplify \eqref{eq:Cauchymatrix} 
and \eqref{eq:barp}  \cite{claytonJMPS2024}. First, elastic shear is assumed small so terms of $O((\gamma^E)^2)$ are negligible. 
Second, thermal expansion in $\tilde{p}$ is omitted (e.g., $A_0 \rightarrow 0$), typical in analysis of shear bands \cite{molinari1986,molinari1987,wright2002,le2018}.
Third, terms of $O(\delta^\xi_0)$, $O(\delta^\phi_0)$ are omitted in the prefactor of \eqref{eq:Cauchymatrix}. 
This implies local volume changes from melting and porosity remain small. 
For ferrous metals studied in \S5 and \S6, such volume changes, both expansive, cannot exceed $\approx 5$\%, which would lead to overestimation of stress, at most, by the same percentage. 
Thus, \eqref{eq:Cauchymatrix} reduces to
\begin{align}
\label{eq:stressred}
& [ {\bm \sigma} ] \approx
\begin{bmatrix}
& -{p}_0 & \mu \gamma^E/J^E & 0 \\
&  \mu \gamma^E/J^E  &  -{p}_0 & 0 \\
& 0 & 0 & -{p}_0 
\end{bmatrix}, 
\quad J^E \approx \underset{\hat{J}^E > 0}{\arg} \{ p_0 = - {B_0}   \frac{ \ln \hat{J}^E}{\hat{J}^E}  [ 1 - {\textstyle{\frac{1}{2}}}  (B'_0 - 2) \ln \hat{J}^E] \}.
\end{align}
Since $p \approx p_0 = \text{constant}$, $J^E(p_0) = \text{constant}$ for $t \geq 0^+$, consistent with \eqref{eq:1D}.

Initial pressure, if tensile, is assumed small enough that no damage is incurred prior to shear loading,
and subsequently, changes in $B$ from shear-induced melting and fracture are ignored in the second of
\eqref{eq:stressred}. 
Transient increases to $J^E$ that would be needed to offset
reduction of $B$ with increasing $\phi$ when $p_0 < 0$ are omitted.
Thus, \eqref{eq:stressred} is most accurate for neutral or compressive states with $p_0 \geq 0$. 
Since $\sigma_{yy} \approx -p_0$, the second of \eqref{eq:mombal1D} is now obeyed unconditionally when $\dot{\upsilon}_y = 0$.
The first of \eqref{eq:mombal1D} is $ \partial \tau / \partial y \approx \partial ( \mu \gamma^E) / \partial y = J^E \rho \dot{\upsilon}_x$.
If $\tau (t)= J^E \sigma_{xy}(t)$ is independent of $y$, then inertial force is negligible.

From \eqref{eq:FE}--\eqref{eq:stressred}, the dissipation contributions from plastic and inelastic
structure deformations of melting and fracture in the energy balance of \eqref{eq:Tdot} and entropy inequality of \eqref{eq:2nd2} are
\begin{align}
\label{eq:pdiss}
& [( {\bf F}^I)^{\rm T} \bar{\bf S} ({\bf F}^I)^{\rm -T}]:{\bf L}^P = \tau \dot{\gamma}^P, \qquad
\tau = J^E \sigma_{xy}  = \mu \gamma^E / J^I \approx \mu \gamma^E, \qquad \bar{p}_0 = J^E p_0 ;
\\
\label{eq:xidiss}
&  [\bar{\bf S} ( {\bf F}^I)^{- {\rm T}}]: \frac{\partial {\bf F}^I }{ \partial \xi } \dot{\xi} =
- \biggr{[} \tau \gamma_0^\xi - \frac{ \{ \tilde{p} -  {\textstyle{\frac{2}{3}}}  \mu (\gamma^E)^2 \} \delta^\xi_0}
{1 + \delta^\xi_0 (1 - \iota^\xi)} \biggr{]}
\frac{\d \iota^\xi}{\d \xi} \dot{\xi}
\approx - [\tau \gamma^\xi_0 - \bar{p}_0 \delta^\xi_0] \frac{\d \iota^\xi}{\d \xi} \dot{\xi} ,
\\
\label{eq:phidiss}
&  [\bar{\bf S} ( {\bf F}^I)^{- {\rm T}}]: \frac{\partial {\bf F}^I }{\partial \phi} \dot{\phi} =
- \biggr{[}\tau \gamma_0^\phi - \frac{ \{\tilde{p} -  {\textstyle{\frac{2}{3}}}  \mu (\gamma^E)^2 \} \delta^\phi_0}
{1 - \delta^\phi_0 (1 - \iota^\phi)} \biggr{]}
\frac{\d \iota^\phi}{\d \phi} \dot{\phi}
\approx -[\tau \gamma^\phi_0 - \bar{p}_0 \delta^\phi_0] \frac{\d \iota^\phi}{\d \phi} \dot{\phi}.
\end{align}
Conjugate forces $\varsigma$ and $\zeta$ of \eqref{eq:varsigma} and \eqref{eq:zeta} become, with
$\Gamma_\xi$ and $\Gamma_\phi$ constants for brevity,
\begin{align}
\label{eq:varsigma2}
 \varsigma & \approx - \mu_0 [{\textstyle{\frac{1}{2}}} \omega ({\rm tr} \tilde {\bf C}^E - 3) + r_0^\xi \bar{R}]  \frac{\d  \iota^\xi }{\d \xi} - 2 A_\xi \xi (1-\xi) ( 1- 2\xi) - h_T [{\theta}/{ \theta_T} -1 ]   \frac{\d  \iota^\xi }{\d \xi} \nonumber
 \\ & \qquad  \qquad  \qquad  \qquad  \qquad  \qquad  \qquad
  - [\tau \gamma^\xi_0 - \bar{p}_0 \delta^\xi_0] \frac{\d \iota^\xi}{\d \xi} 
   + 2 l_\xi \Gamma_\xi \nabla^2_0  \xi ,
   \\
\label{eq:zeta2}
 \zeta & \approx - \mu_0 [{\textstyle{\frac{1}{2}}} ({\rm tr} \tilde {\bf C}^E - 3) ] \iota^\xi \frac{\d {\omega}}{\d \phi} -  E_C \frac{\d f_\phi}{\d \phi}
 -[\tau \gamma^\phi_0 - \bar{p}_0 \delta^\phi_0] \frac{\d \iota^\phi}{\d \phi}
  + 2 l_\phi \Gamma_\phi \nabla^2_0 \phi .
\end{align}
Temperature rate \eqref{eq:Tdot} becomes,
with generally transient Taylor-Quinney factor  
$\beta(y(Y,t),t)$  \cite{claytonJMPS2005,lieou2021} , $\theta = \theta(y(Y,t),t)$, $\dot{J}^E(p_0) = 0$, $A_0 = 0$, and
$\kappa = \text{constant}$ for brevity and simplicity,
\begin{align}
\label{eq:1stbeta}
& c_V \dot{\theta} \approx \beta \tau \dot{\gamma}^P +  \biggr{[ } \varsigma + h_T \frac{ \theta}{ \theta_T} \frac{\d \iota^\xi }{ \d \xi} \biggr{]} \dot{\xi}  + \zeta \dot{\phi} 
+ \kappa \frac{\partial^2 \theta}{\partial Y^2}, 
\quad \beta = \frac{\tau \dot{\gamma}^P - \mu_0 [1-(1-\iota^\xi)r_0^\xi] (\frac{\partial \bar{R}}{ \partial \chi}) \dot{\chi} }{ \tau \dot{\gamma}^P}.
\end{align}
Here in \eqref{eq:1stbeta}, $\beta$ accounts for stored energy of cold work but not energies of fracture and melting.

 Let an arbitrary rigid body rotation be ${\bf R}_0$. Strain energy $W$ and stress ${\bm \sigma}$ are unaffected by transformations of the form ${\bf F}^E \rightarrow {\bf F}^E {\bf R}_0^{\rm T}$ with ${\bf F}^I {\bf F}^P \rightarrow {\bf R}_0 {\bf F}^I {\bf F}^P$. Different choices of ${\bf R}_0$ can affect partitioning of dissipation among inelastic deformation mechanisms. Currently setting ${\bf R}_0 = {\bf 1}$, motivated by physics specific to Fig.~\ref{fig1} and metal plasticity, is not arbitrary \cite{bammann1987,claytonNCM2011}.

\subsection{Approximations and reduced governing equations}
The following additional assumptions are implemented later to facilitate analysis, in conjunction with prior assumptions that $A_0 \rightarrow 0$, $\delta^\xi_0$ and $\delta^\phi_0$ are small compared to unity, and $B \approx B_0 = \rm{constant}$:
\begin{enumerate} \setlength\itemsep{-0.5em} 
\item Inertia is omitted ($\dot{\upsilon}_x \rightarrow 0$), as in Refs.~\cite{molinari1986,molinari1987,claytonJMPS2024};
\item The shearing response is idealized as rigid-viscoplastic, as in 
Refs.~\cite{molinari1986,molinari1987,claytonJMPS2024,wright2002};
\item Gradient regularization is omitted ($l_\phi \rightarrow 0, l_\xi \rightarrow 0$)~\cite{levitas2011,claytonPRE2024}, and $\theta_I \rightarrow \theta_T \Rightarrow A_\xi \rightarrow 0$;
\item Heat conduction is omitted ($\kappa \rightarrow 0$), as in Refs.~\cite{molinari1986,molinari1987,claytonJMPS2024};
\item Taylor-Quinney factor is constant ($\beta \dot{\gamma}^P \rightarrow \beta_0 \dot{\gamma}, \beta_0 = {\rm constant} \in (0,1]$), as in Refs.~\cite{molinari1986,molinari1987,claytonJMPS2024}.
\end{enumerate}

From the first assumption, the lone non-trivial equilibrium equation in the first of \eqref{eq:mombal1D} becomes
\begin{align}
\label{eq:linmomcons}
&\frac{\partial \tau(y,t) }{\partial y } = 0
\quad
 \Rightarrow \quad
\frac{\partial \tau(y(Y,t),t)}{\partial Y} = F_{yY} \frac{\partial \tau}{ \partial y} = 0 \quad
 \Rightarrow \quad 
\tau_A = \tau(Y_A,t) = \tau(Y_B,t) = \tau_B.
\end{align}
 Coordinates of any two material points the slab with shear stresses $\tau_A$, $\tau_B$ are written $Y_A$, $Y_B$. Recall $\tau = J^E \sigma_{xy} \approx \mu \gamma^E$ and from \eqref{eq:FE} and \eqref{eq:FP} that $F_{yY}(Y,t) = (J^E)^{1/3}J^I(Y,t) > 0$.
 Physical suitability of the first and fourth assumptions (i.e., omission of inertia and conduction)
 for materials and loading conditions of present interest, namely steels deformed in torsion at
 rates $\dot{\bar{\gamma}} \in  [10^3$/s$,10^4$/s$]$ is
 discussed in Refs.~\cite{molinari1986,molinari1987,claytonJMPS2024,wright1994,wright2002}.
 Limitations of these assumptions are reconsidered in \S5.2 and \S7.
 Acceleration is more important at much higher rates, conduction at much lower rates.
 
 From the second assumption above, rigid viscoplasticity sets $\tau/\mu \rightarrow 0$ but $\tau$ can remain finite.
Consistent with this limit, $\gamma^E \rightarrow 0$,
$\tau / \mu_0 \rightarrow 0$, $\tau \rightarrow 0$ as $\iota^\xi \omega \rightarrow 0$, 
 and quadratic deviatoric strain energy $\mu ( {\rm tr} \tilde{\bf C}^E - 3) \rightarrow 0$.
With $J^I$ on the order of unity, the shear strain and shear strain rate, with explicit viscoplastic and inelastic structural deformations, reduce to 
\begin{align}
\label{eq:redmodela}
& \gamma \approx \gamma^P + \gamma^\xi_0 (1-\iota^\xi) + \gamma^\phi_0 (1 - \iota^\phi)   \geq 0, \qquad \dot{\gamma} \approx \dot{\gamma}^P  - \gamma_0^\phi \frac{ \d \iota^\phi}{\d \phi} \dot{\phi} - \gamma_0^\xi \frac{ \d \iota^\xi}{\d \xi} \dot{\xi} \geq 0 .
\end{align}
If $\gamma^\xi + \gamma^\phi < 0$, shearing from slip must exceed that from fracture and melting. If 
$\gamma^\xi + \gamma^\phi > 0$, then $\gamma^P < 0$ is not impossible. 

The third assumption discards some energetic contributions of boundary layers between
solid and liquid and between undamaged and fully fractured material.
While this assumption neglects certain physics resolved by phase-field theories of
gradient type \cite{claytonJMPS2021,claytonIJF2014,levitas2011b,hwang2014}, it allows for modeling of spatial discontinuities in order parameters, 
and thus does not implicitly forbid jumps in shear strain $\gamma$ and temperature $\theta$ needed for the locally infinite-$\gamma$ description of adiabatic shear bands \cite{molinari1986,molinari1987,molinari1988,claytonJMPS2024} invoked in \S4--\S6.
Were the gradient energy contribution from $\xi$ included, a sharp spatial change or jump
in $\xi$ would be impossible due to the nearly infinite or infinite energetic barrier from $\Lambda$ of \eqref{eq:graden}.
A similar remark applies for the gradient energy contribution from $\phi$ in \eqref{eq:graden}.
Since primary driving forces for local increases in $\xi$ and $\phi$ are later identified with
$\theta$ and $\gamma$, respectively, the same barriers could restrict large increases in
$\theta$ and $\gamma$ to zones or bands of finite width. A jump to an extreme
value of $\theta$ or $\gamma$ for material in an infinitesimal-width band relative to its surroundings
would engender a discontinuity in $\xi$ or $\phi$ that is inadmissible in a gradient-regularized
theory. Conductivity $\kappa > 0$ would also restrict the minimum width of a shear band to
avoid infinite contributions to dissipation in \eqref{eq:2nd2} and temperature rate in \eqref{eq:1stbeta}, as
temperature gradients approach infinite magnitudes.
Combining the third assumption with the second,
\begin{align}
\label{eq:varsigmazeta3}
 \varsigma & \approx - \biggr{[}  \mu_0 r_0^\xi \bar{R}  
 + h_T \{ {\theta}/{ \theta_T} -1 \} 
  + \tau \gamma^\xi_0 - \bar{p}_0 \delta^\xi_0  \biggr{]}  \frac{\d \iota^\xi}{\d \xi},
\quad
 \zeta  \approx  -  E_C \frac{\d f_\phi}{\d \phi}
 -[\tau \gamma^\phi_0 - \bar{p}_0 \delta^\phi_0] \frac{\d \iota^\phi}{\d \phi} .
\end{align}

Finally, from the fourth and fifth assumptions, the first inequality in \eqref{eq:2nd2}, 
the first equality in \eqref{eq:1stbeta}, and the total
cumulative plastic work $W^P$ are approximated, respectively, by
\begin{align}
\label{eq:2nd3}
&  \beta_0 \tau \dot{\gamma} + \varsigma \dot{\xi} + \zeta \dot{\phi} \gtrsim 0,
\quad c_V \dot{\theta} \approx \beta_0 \tau \dot{\gamma} 
+  \biggr{[ } \varsigma + h_T \frac{ \theta}{ \theta_T} \frac{\d \iota^\xi }{ \d \xi} \biggr{]} \dot{\xi}
+ \zeta \dot{\phi},
\quad
W^P \approx \int_0^t \tau \dot{\gamma} \, \d \hat{t} = \int_0^\gamma \tau \, \d \hat{\gamma}.
\end{align}
From \eqref{eq:redmodela} with $\dot{\gamma}^E = 0$, if $\gamma^\xi_0 \dot{\xi} \neq 0$  or $\gamma^\phi_0 \dot{\phi} \neq 0$, then
$\dot{\gamma}^P \neq \dot{\gamma}$.
In that case, the fifth assumption implies $\beta  = \beta(t) = \beta_0 (1 + \dot{\gamma}^P/\dot{\gamma}^I)$ can be transient if $\beta_0 = \text{constant}$,
the latter implicitly affected by dissipative contributions from inelastic structural shears. In 
\S5 and \S6, $\gamma^\xi_0 = \gamma^\phi_0 = 0 \Rightarrow \beta \rightarrow \beta_0$.

Models for plasticity, ductile damage, and melting following in \S3.3, \S3.4, and \S3.5 are specialized
to the dynamic, adiabatic, pressurized simple shear problem of \S3.1, simplifying assumptions of \S3.2, and relatively strong, ductile polycrystalline metals like structural steels. Different models would be needed for other loading regimes and material classes (e.g., quasi-statics, brittle solids).

\subsection{Viscoplasticity}
A power-law viscoplastic flow rule \cite{fressengeas1987,molinari1986,molinari1987,claytonJMPS2024} relates $\dot{\gamma}$ to Kirchhoff shear stress $\tau = J^E \sigma_{xy} \geq 0$:
\begin{align}
\label{eq:flowrule}
& \dot{\gamma} = \dot{\gamma_0} \, ( \tau / g)^{1/m}, \qquad
g = g(\xi,\phi,\chi,\theta; p_0) = g_Y (\xi, \phi, p_0) h (\chi) \lambda (\theta).
\end{align}
Denoted by $\dot{\gamma}_0 = \text{constant} \geq 0$ and $m = \text{constant} > 0$ are a reference strain rate and rate sensitivity. Flow stress $g \geq 0$ consists of pressure-, phase-, and damage-dependent static yield stress $g_Y \geq 0$, strain hardening (or softening) function $h \geq 0$, and thermal softening (or hardening) function $\lambda \geq 0$. 

Internal state variable change $\chi(t) - \chi_0$ is associated with strain hardening, most often linked to increases in dislocation density \cite{kocks1975,claytonNCM2011}. For monotonically increasing $\gamma(t)$, 
the simplest realistic representation of $\chi$ associated with dislocations is,  with initial value $\chi_0$ close to unity, 
\begin{align}
\label{eq:chi}
\chi(y,t) = \chi_0(Y(y,t)) + \gamma(y,t), \quad \chi_0 \approx 1;
\quad
\dot{\chi} = \dot{\gamma} \, \Rightarrow \,
\bar{R} \approx \frac{1-\beta_0}{\mu_0} \int_0^\gamma \frac{\tau \, \d \hat{\gamma}}
{1-(1-\iota^\xi)r^\xi_0} + \bar{R}_0.
\end{align}
In the rightmost expression, obtained from \eqref{eq:1stbeta} and the fifth assumption of \S3.2 with
$\beta_0 \approx \beta$, 
$\bar{R}_0(Y) \geq 0$ is an initial condition related to stored energy from initial dislocation density at $Y$. 

Fractured or molten material is assumed to have degraded strength. Classical dislocation plasticity theory
\cite{kocks1975,claytonNCM2011} presumes $g_Y$ is proportional to shear modulus $\mu$. Thus
$g_Y$ is interpolated akin to $\mu$ in \eqref{eq:moduli} via functions $\iota^\xi(\xi)$ and $\omega(\phi)$.  As in
 Refs.~\cite{fressengeas1987,molinari1986,molinari1987,claytonJMPS2024}, $h$ and $\lambda$ are power-law forms:
\begin{align}
\label{eq:taulaw}
& g_Y(\xi,\phi,p_0) = g_0(p_0) \iota^\xi(\xi) \omega(\phi), \qquad h(\chi(\gamma)) = \chi_0 (1 + \gamma / \gamma_0)^n, \qquad \lambda(\theta) =  ( \theta / \theta_0)^\nu.
\end{align}
Here, $g_0(p_0) > 0$ is strength of the solid with pressure scaling, while $\gamma_0$, $n$, and $\nu$ are dimensionless constants. 
Strength depends linearly on $p_0$ via the pressure derivative of shear modulus, $\mu'$ \cite{claytonJMPS2024}:
\begin{align}
\label{eq:gyp}
g_0(p_0) = [1 + (\mu'/\mu_0) p_0] g_0(0); \qquad \mu' = (\partial \mu / \partial p)_{\theta = \theta_0,p= 0}.
\end{align}
To avoid unnecessary complexity that would be eliminated by the rigid-viscoplastic assumption, pressure scaling of $\mu$ was omitted in elastic strain energy \eqref{eq:W}. 
Flow stress $\tau$ is, inverting \eqref{eq:flowrule}, 
\begin{align}
\label{eq:flowstress}
 \tau(\gamma(y,t), \dot{\gamma}(y,t), \xi(y,t),\phi(y,t),\theta(y,t); \chi_0(y),p_0)   = \chi_0   g_0(p_0) 
 \iota^\xi(\xi) \omega(\phi) 
 \biggr{(} 1 + \frac{\gamma}{ \gamma_0} \biggr{)}^n  \biggr{(} \frac{\theta }{ \theta_0} \biggr{)}^\nu \biggr{(} \frac{\dot{\gamma}}{\dot{\gamma}_0} \biggr{)}^m.
\end{align}
In metals, the strain hardening exponent $n$ is usually positive, and thermal softening exponent $\nu$ is usually
negative. These conventions are not enforced a priori as restrictions in the theory, however.

Accommodating usual degradation functions from continuum-damage \cite{claytonNCM2011,claytonNEIM2019,claytonIJSS2015,claytonCMT2022} and phase-field mechanics \cite{claytonIJF2014,claytonJMPS2021} the function $\omega(\phi)$ first introduced in \eqref{eq:moduli} and
presently entering \eqref{eq:taulaw} is
\begin{equation}
\label{eq:omega}
\omega(\phi) = \omega_1 + (1-\omega_1)(1-\phi)^k, \qquad \d \omega / \d \phi = -k (1-\omega_1) (1-\phi)^{k-1}.
\end{equation}
 Material constants are $k \geq 1$ and $\omega_1 \in [0,1]$, recalling the latter for remnant strength at $\phi = 1$.
 
Initial deviation $\delta \chi_0 (Y) = 1 -\chi_0 (Y)$, if positive as assigned by convention in \S4--\S6, accounts for strength defects.
These could be initial micro-cracks, pores \cite{vishnu2022}, texture variations, weak interfaces at grain boundaries \cite{claytonJMPS2005,claytonIJSS2005}, inclusions, or another
microstructure feature that reduces strength.
Deviation can be related to time-independent ``damage'' parameter $\tilde{\phi}_0$ with linear degradation
function $\tilde{\omega}_0$:
\begin{align}
\label{eq:hatphi0}
 \delta \chi_0 (Y) = 1 - \chi_0(Y) = \tilde{\phi}_0(Y), \qquad \tilde{\omega}_0 (\tilde{\phi}_0)  = 1 - \tilde{\phi}_0 = \chi_0.
\end{align}
This is akin to \eqref{eq:omega} with $\phi \rightarrow \tilde{\phi}_0$, $k \rightarrow 1$, and $\omega_1 \rightarrow 0$.
In the present theory, it is advantageous to assign distinct values to $\tilde{\phi}_0$ and $\phi_0 = \phi(t = 0)$,
whereby $\phi_0 = 0$ is used as the standard initial condition.
This allows $\tilde{\phi}_0$ and $\phi$ to describe different weakening mechanisms (e.g., grains locally oriented favorably for easy slip, unrelated to ductile fracture).

If the solid has mild rate sensitivity (i.e., $m$ small versus unity), 
an assumption used in some, but not all, steps of prior analyses \cite{molinari1986,molinari1987,claytonJMPS2024} replaces local strain rate in \eqref{eq:flowstress} with average $\dot{\bar{\gamma}}$: 
\begin{align}
\label{eq:flowstress2}
 & \tau(\gamma,\xi,\phi,\theta; \dot{\bar{\gamma}}; \chi_0, p_0) \approx \chi_0 g_0(p_0)  \iota^\xi(\xi) \omega(\phi) 
 \biggr{(} 1 + \frac{\gamma}{ \gamma_0} \biggr{)}^n  \biggr{(} \frac{\theta }{ \theta_0} \biggr{)}^\nu \biggr{(} \frac{\dot{\bar{\gamma}}}{\dot{\gamma}_0} \biggr{)}^m, 
\quad \dot{\bar{\gamma}} = \frac{\upsilon_0}{ h_0}.
\end{align}
More accurate \eqref{eq:flowstress} is used in the momentum balance. Approximation \eqref{eq:flowstress2} is used to estimate dissipative contributions to temperature \cite{molinari1986,molinari1987} and structure kinetics \cite{claytonJMPS2024} if $\dot{\gamma}(y,t)$ is unknown.

\subsection{Shear fracture}
Classical, elastic-brittle phase-field fracture models \cite{karma2001,claytonIJF2014,miehe2015a,claytonIJSS2019} are inadequate for describing ductile failure in adiabatic shear bands because they
do not account for coupling of plastic deformation to effective fracture toughness and fracture kinetics.
Rather, similar to Refs.~\cite{miehe2015b,mcauliffe2015,mcauliffe2016,choo2018,na2018,wang2020,samaniego2021,zeng2022},
here some fraction of plastic working is used as a driving force for shear fracture.

Let $\alpha_0 \in [0,\chi_0]$ be
a dimensionless constant quantifying this fraction. 
Degradation initiates when a threshold level of plastic work $W^\phi$ has been attained
\cite{dolinski2015,dolinski2015b,wang2020,samaniego2021,miehe2015b,borden2016}. 
Set $W^\phi = \chi_0 W^{\phi}_0 [1 + (\mu'/\mu_0) p_0]$,
where  $W^{\phi}_0 = {\rm constant} \geq 0$. Accordingly, initial defects and tensile pressure are assumed to linearly reduce threshold $W^\phi$, analogously to static yield-strength terms $h$ and $g_Y$ in \eqref{eq:flowstress} and \eqref{eq:gyp}.
Ensuring net non-negative dissipation from plastic work and ductile fracture, the following kinetic law for the order parameter rate, $\dot{\phi}$, for dynamic shear degradation is postulated:
\begin{align}
\label{eq:phidot}
&\zeta \dot{\phi} = - [{\alpha_0 \beta_0}/{\chi_0}] \tau \dot{\gamma} \, {\mathsf{H}}(W^P - W^\phi) 
{\mathsf H} (1-\phi)  \nonumber
 \\ & \quad \Rightarrow 
\quad \beta_0 \tau \dot{\gamma} + \zeta \dot{\phi} =
\beta_0 \tau [ 1 - ( \alpha_0 / \chi_0) {\mathsf{H}}(W^P - W^\phi) {\mathsf H} (1-\phi) ] \dot{\gamma} \geq 0.
\end{align}
Heaviside function ${\mathsf H}(1-\phi)$ limits the solution to physical domain $\phi \in [0,1]$.
Substituting \eqref{eq:varsigmazeta3} into the first of \eqref{eq:phidot} and enforcing $\d \phi \geq 0$ gives a nonlinear differential equation at each $y \in [0,h]$:
\begin{align}
\label{eq:phidot2}
  \biggr{ [} \langle E_C \frac{ \d f_\phi}{\d \phi} 
  +  \{ \tau(\phi, \cdot) \gamma^\phi_0 - \bar{p}_0 \delta^\phi_0 \} \frac{\d \iota^\phi (\phi)} {\d \phi}
   \rangle   \biggr{]} {\d \phi} 
  = 
 \biggr{ [ } 
 \frac{ \alpha_0 \beta_0}{\chi_0} \tau(\phi,\cdot) {\mathsf{H}}(W^P - W^\phi) {\mathsf{H}}(1-\phi) 
 \biggr{]}
  { \d \gamma}, 
\end{align}
where $\langle (\cdot) \rangle = \frac{1}{2}[(\cdot) + |(\cdot)|]$.
The right side of \eqref{eq:phidot2} is non-negative since $\tau \geq 0$ and $\d \gamma \geq 0$.
The $\langle (\cdot) \rangle$ operation on the left ensures $\d \phi \geq 0$ (i.e., irreversible damage), and
$\phi \rightarrow 1$ by the ${\mathsf H}(1-\phi)$ factor on the right when resistance $-\zeta$ within $\langle (\cdot) \rangle$ on the left is non-positive.
This equation must usually be integrated numerically for $\phi$ using \eqref{eq:flowstress} along with
evolution equations for $\xi$ and $\theta$.

To demonstrate model features, apply \eqref{eq:flowstress2} in each of \eqref{eq:2nd3} \cite{molinari1986,molinari1987,claytonJMPS2024} and assume $\gamma_0^\phi = 0$, meaning no explicit inelastic shear strain from fracture.
For monotonically increasing $\gamma(y,t)$, assume functional forms $\theta(y,t) = \theta(\gamma(y,t),y)$ and 
$\xi(y,t) = \xi(\gamma(y,t),y)$ exist.
Denote by $\gamma_f (y)$ the value of $\gamma(y,t)$ at $t$ when fracture initiates, as $W^P \rightarrow W^\phi$.
Then \eqref{eq:phidot2} is separable and can be integrated for $\phi(\gamma) \in [0,1]$
at fixed $\bar{p}_0$, $\chi_0$, and $y$:
\begin{align}
\label{eq:phisol1}
\int_0^\phi  
 \langle \frac{E_C }{ \omega (\hat{\phi}) } \frac{ \d f_\phi}{\d \hat{\phi}} 
- \frac{\bar{p}_0 \delta^\phi_0}{\omega (\hat{\phi}) } \frac{ \d \iota^\phi}{\d \hat{\phi}} \rangle  
{\d \hat{\phi} }=
\alpha_0 \beta_0  g_0
\biggr{(}  \frac{\dot{\bar{\gamma}}}{\dot{\gamma}_0} \biggr{)}^m
 \int_{\gamma_f}^\gamma 
 \iota^\xi(\xi(\hat{\gamma})) 
 \biggr{(} 1 + \frac{ \hat{\gamma} }{ \gamma_0} \biggr{)}^n  \biggr{(} \frac{\theta (\hat{\gamma}) }{ \theta_0} \biggr{)}^\nu 
  \d \hat{\gamma} .
\end{align}
Now take $f_\phi = \phi$, $k = 1$ in \eqref{eq:omega}, $\iota^\phi = 1 - \phi$,
and assume $E_C + \bar{p}_0 \delta^\phi_0 > 0$. Then \eqref{eq:phisol1} produces
\begin{align}
\label{eq:phisol2}
\phi(\gamma) = \min \left[ \frac{1 - \exp \{ -A_f W^P_f (\gamma) \} }{1-\omega_1}, 1\right]; 
\quad A_f = \frac{\alpha_0 \beta_0 (1 - \omega_1)}{\chi_0 (E_C + \bar{p}_0 \delta^\phi_0)},
\quad W^P_f =  \int_{\gamma_f}^\gamma \frac{\tau}{\omega} \d \hat{\gamma}.
\end{align}
As plastic work $W^P_f \rightarrow \infty$, $\phi \rightarrow 1$. When $\gamma(y,t) \leq \gamma_f (y)$, the trivial solution is $\phi(y,t) = 0$,
and $\d \phi = 0$ when $\tau = 0$. Resistance to fracture at a given value of $\gamma$ is afforded by $E_C$; large  $\alpha_0$, $\beta_0$ and defects $\delta \chi_0  = 1-\chi_0 \geq 0 $ promote fracture.
If $E_C \leq  \bar{p}_0 \delta^\phi_0$, with ductile damage expansive (i.e., $\delta^\phi_0 > 0$) tensile pressure $- \bar{p}_0 \geq E_C / \delta^\phi_0$ causes fracture as soon as $W^P > W^\phi$ via $\phi \rightarrow 1$ in \eqref{eq:phidot2}.

\subsection{Melting}
To ensure non-negative dissipation from melting or solidification, the Allen-Cahn or Ginzburg-Landau equation \cite{levitas2011b,hwang2014,hwang2016} is often invoked for kinetics of ${\xi}$, especially when modeling
boundary layers of partially molten material. Here, since gradient regularization is omitted and solidification is
less relevant, a linear
relaxation model for phase transformations \cite{boettger1997,claytonCMT2022,claytonZAMM2024}, also obeying non-negative dissipation, is instead adapted for melting. Letting $t_R^\xi \geq 0$ be a relaxation time constant, 
\begin{align}
\label{eq:xidot}
t_R^\xi \dot{\xi} = \begin{cases}
 \quad \! \! \langle \bar{\xi}- \xi \rangle, \qquad   & (\varsigma > 0),
\\    -  \langle  \xi -  \bar{\xi} \rangle, \qquad & (\varsigma \leq 0)
\end{cases}
\quad \Rightarrow \quad \varsigma \dot{\xi} \geq 0.
\end{align}
The metastable melt fraction is $\bar{\xi}$. With $\alpha_\xi \geq 0$ and $\beta_\xi > 0$ kinetic barriers, the following ordinary differential equation (ODE) is posited for melting but not solidification, akin to Refs.~\cite{boettger1997,claytonCMT2022,claytonZAMM2024}:
\begin{align}
\label{eq:dxi}
\d \bar{\xi} =  - \iota^\prime \d \biggr{[}  \frac{ \langle - \varsigma / \iota^\prime -\alpha_\xi \rangle }{ \beta_\xi } \biggr{]} { \mathsf H} (1-\bar{\xi}); \qquad \iota^{\prime} (\bar{\xi}) = \frac{\d \iota^\xi (\bar{\xi})}{ \d \bar{\xi}} \leq 0, \quad \iota^\prime(1) = 0.
\end{align}
The Heaviside function restricts the solution to $\bar{\xi} \leq 1$. 
This differential equation must generally be solved numerically for $\bar{\xi}$ since $\varsigma$ in \eqref{eq:varsigmazeta3} depends implicitly on $\bar{\xi}$.
 Dividing by $\iota^\prime$ and integrating,
\begin{align}
\label{eq:xisol1}
\int_0^{\bar{\xi}} \frac{ \d {\xi}}{ \iota^\prime}   
=   - \frac{1}{\beta_\xi} \langle \mu_0  r_0^\xi \bar{R} 
 + \frac{h_T}{\theta_T} \{ \theta - \theta_T \} 
 + \{ \tau \gamma^\xi_0 - \bar{p}_0 \delta^\xi_0 \} - \alpha_\xi \rangle, \qquad (\bar{\xi} \leq 1).
\end{align}

Consider another analytical example to illustrate concepts and model features. When $t_R^\xi \dot{\bar{\gamma}} \ll 1$, the left side of \eqref{eq:xidot} is assumed to vanish, whereby $\xi \approx \bar{\xi}$. 
Assume such conditions hold, meaning melting completes at time scales fast relative to the loading time.
Continuing to assume forms $\theta(\gamma(y,t),y)$ and $\phi(\gamma(y,t),y)$ akin to those in \eqref{eq:phisol1} exist, and 
choosing, for example, 
\begin{align}
\label{eq:interpxi}
\iota^\xi(\xi)  = \iota^\xi_1 + (1-\iota^\xi_1)(1-\xi)^2,
\quad
\iota^\prime = {\d \iota^\xi}/{\d \xi} = - 2 (1-\iota^\xi_1) (1 -\xi) \in [ - 2 (1-\iota^\xi_1), 0] ,
\end{align}
 \eqref{eq:xisol1} can be solved analytically as
\begin{align}
\label{eq:xisol2}
\xi(\gamma) = 1 - \exp \biggr{[} - \frac{2 (1-\iota_1^\xi)}{\beta_\xi} \langle \Xi (\gamma) \rangle \biggr{]}, \quad
\Xi = \frac{h_T}{\theta_T} \{ \theta - \theta_T \} +  \mu_0 r_0^\xi \bar{R} 
 +  \tau \gamma^\xi_0 - \bar{p}_0 \delta^\xi_0  - \alpha_\xi.
\end{align}
If all terms except the first in $\Xi$ vanish, then the local melt fraction $\xi = 0$ for $\theta \leq \theta_T$ and $\xi$ increases with $\theta$ for $\theta > \theta_T$. Solidification is possible (i.e., $\d \xi < 0$) in that case if $\theta$ subsequently decreases.
However, \eqref{eq:dxi} with $\bar{\xi} \rightarrow \xi$ ensures non-negative dissipation $ \varsigma \d \xi \geq 0$ only when
$\d \xi \geq 0$, with $\varsigma < 0 \Rightarrow \d \xi = 0$.  
The case $\d \xi < 0$ when $\varsigma > 0$ is not ruled out.
In calculations, non-negative dissipation
can be enforced by the constraint $ \d \xi = \langle \d \bar{\xi} \rangle$.
Solidification is accordingly eliminated.
A different ODE involving some sign changes to \eqref{eq:dxi} is needed to model dissipative, metastable reverse transitions (i.e., solidification)
\cite{boettger1997,claytonCMT2022,claytonZAMM2024}.

\subsection{Temperature}
Substituting \eqref{eq:varsigmazeta3}, \eqref{eq:flowstress2}, and \eqref{eq:phidot} into \eqref{eq:2nd3}, for $\phi \leq 1$ and $\gamma^\xi_0 = 0$ as invoked in \S4--\S6, gives
\begin{align}
\label{eq:Tdot3}
\dot{\theta} & =  \frac{\beta_0}{c_V} [1 - \frac{\alpha_0}{ \chi_0} {\mathsf H}(W^P - W^\phi )] 
\chi_0 g_0 \omega  \iota^\xi  \biggr{(} 1 + \frac{\gamma}{ \gamma_0} \biggr{)}^n  \biggr{(} \frac{\theta }{ \theta_0} \biggr{)}^\nu \biggr{(} \frac{\dot{\bar{\gamma}}}{\dot{\gamma}_0} \biggr{)}^m \! \! \dot{\gamma}
-  \frac{  [\mu_0 r_0^\xi \bar{R}  
 - h_T  
 - \bar{p}_0 \delta^\xi_0 ] } {c_V}
  \frac{\d \iota^\xi} {\d \xi} \dot{\xi}.
\end{align}
This nonlinear differential equation must generally be solved numerically for $\theta(\gamma,y)$ at each $y \in [0,h]$, in conjunction with \eqref{eq:phidot2} for $\phi (\gamma,y)$, \eqref{eq:xidot} for $\xi(\gamma,y)$, and \eqref{eq:chi}
for $\bar{R}(\gamma,y)$. 
If functional forms $\phi(\gamma(y,t),y)$, $\xi(\gamma(y,t),y)$, and $\bar{R}(\gamma(y,t),y)$ exist, then at fixed $y$
\eqref{eq:Tdot3} can be written
\begin{align}
\label{eq:Tdot4}
\frac{\d {\theta}}{ \d \gamma} & =  \frac{\beta_0}{c_V} [1 - \frac{\alpha_0}{ \chi_0} {\mathsf H}(W^P - W^\phi )] 
\chi_0 g_0 \omega  \iota^\xi  \biggr{(} 1 + \frac{\gamma}{ \gamma_0} \biggr{)}^n  \biggr{(} \frac{\theta }{ \theta_0} \biggr{)}^\nu \biggr{(} \frac{\dot{\bar{\gamma}}}{\dot{\gamma}_0} \biggr{)}^m \! \! \! \! \!
-  \frac{ [  \mu_0  r_0^\xi \bar{R}  
 - h_T  
 - \bar{p}_0 \delta^\xi_0  ] }
 {c_V}
 \frac{\d \iota^\xi}{\d \xi} \frac{ \d {\xi}}{\d \gamma} .
\end{align}
This equation likewise must be solved numerically. But if melting never occurs, $\xi = 0 \Rightarrow \iota^\xi = 1$, and, given function $\phi(\gamma)$ at $y$, \eqref{eq:Tdot4} can be separated and integrated from $\theta(y,0) = \theta_i (y)$ as
\begin{align}
\label{eq:thetanomelt}
&   {\theta}^{-\nu} \d {\theta}  = \biggr{ \{} \frac{\beta_0}{c_V \theta_0^\nu} [1 - (\alpha_0/ \chi_0) {\mathsf H}(W^P - W^\phi )] 
\chi_0 g_0 \omega \biggr{(} 1 + \frac{\gamma}{ \gamma_0} \biggr{)}^n  \biggr{(} \frac{\dot{\bar{\gamma}}}{\dot{\gamma}_0} \biggr{)}^m \biggr{ \}} \d \gamma \quad \Rightarrow \quad \nonumber
\\ 
& \theta (\gamma)  = 
{\pmb{ \biggr{[} }} \theta_i^{1-\nu} + \frac{(1-\nu)  \beta_0 \chi_0 g_0  \gamma_0} {(1+n) c_V \theta_0^\nu} 
 \biggr{(}  \frac{\dot{\bar{\gamma}}} {\dot{\gamma}_0} \biggr{)}^m
\biggr{\{} \biggr{(}1 + \frac{\gamma_f}{ \gamma_0} \biggr{)}^{1+n}-1 \biggr{\}} 
\nonumber \\
& \qquad \, \, \,
+ \frac{(1-\nu) (\chi_0 - \alpha_0) \beta_0  g_0 }{c_V \theta_0^\nu} \biggr{(}  \frac{\dot{\bar{\gamma}}} {\dot{\gamma}_0} \biggr{)}^m \int_{\gamma_f}^\gamma  \omega(\phi (\hat{\gamma})) \biggr{(} 1 + \frac{\hat{\gamma}}{ \gamma_0} \biggr{)}^n \d \hat{\gamma }{\pmb{ \biggr{]}} }^{\textstyle{\frac{1}{1-\nu}}},
\quad (\text{if no melting}).
\end{align}
The solution for $\theta(\gamma)$ in \eqref{eq:thetanomelt} presumes fracture occurs during the strain history,
that is, $\gamma \geq \gamma_f$. If $\gamma < \gamma_f$, \eqref{eq:thetanomelt} holds with
$\gamma_f \rightarrow \gamma$ such that the integral embedded on the right involving $\omega$ vanishes.
In the latter case, the analytical solution matches Ref.~\cite{claytonJMPS2024} in the absence of phase transformations.

\section{Localization and numerical methods \label{sec4}}

\subsection{Failure modes}

Three different shear failure criteria are defined:
\begin{itemize}  \setlength\itemsep{-0.1em} 
\item \textbf{Shear banding}. This is the $L_\infty$ localization definition of Molinari and Clifton \cite{molinari1986,molinari1987,molinari1988} also used in Ref.~\cite{claytonJMPS2024}.
With reference to the last of \eqref{eq:linmomcons}, failure by localization of shear strain $\gamma(Y,t)$ occurs at material point $B$ with $Y = Y_B$ and
time $t_c$ when $ \gamma(Y_B,t) / \gamma(Y_A,t) = \gamma_B(t) / \gamma_A(t) \rightarrow \infty$ with increasing time $t \geq t_c$ for every point  $A$  with $Y_A \neq Y_B$.
If $\dot{\gamma}$ remains bounded, $\tau \rightarrow 0$ as $\gamma_B \rightarrow \infty$, meaning shear stress vanishes at some time $t > t_c$.
\item \textbf{Shear fracture}. As in phase-field and continuum damage mechanics, shear fracture occurs when
order parameter $\phi \rightarrow 1$. In the present analysis, shear fracture will generally occur earliest at a point $Y_B$
due to an initial defect. If $\omega_1 = 0$ in \eqref{eq:omega}, then $\tau \rightarrow 0$ upon shear fracture.
If $\omega_1 > 0$, some (small) fraction of strength can be maintained, depending on whether shear banding or melting take place concurrently.
\item \textbf{Melting}. Failure by melting occurs when order parameter $\xi \rightarrow 1$. In the present analysis,
melting will tend to occur first where temperature rise is largest, which correlates with high-strain regions triggered by initial defects (e.g., at point $Y_B$). In \eqref{eq:interpxi}, $\tau \rightarrow 0$ as $\xi \rightarrow 1$ if $\iota^\xi_1 = 0$. But if $\iota^\xi_1 > 0$, then some fraction of strength can be maintained depending on whether shear banding or shear fracture occur simultaneously.
\end{itemize}
Shear band failure, by definition, requires infinite strain at $Y_B$. Shear fracture and melting may or may not ensue infinite strain, depending on assumptions and parameters entering their governing equations.
For example, $\gamma_B \rightarrow \infty$ is required for $\phi \rightarrow 1$ in \eqref{eq:phisol2} if
$\omega_1 = 0$, but not if $\omega_1 > 0$.
The model in \eqref{eq:xisol2} requires $\gamma_B \rightarrow \infty$ to approach infinite $\theta$ or $\bar{R}$  for $\xi \rightarrow 1$. If infinite strains are not required, fracture or melt failure can precede shear band failure. Depending on constitutive parameters for viscoplasticity, fracture, and melting, one or more failure modes may be impossible.
Using a different theory, phase-field simulations \cite{arriaga2017} suggested a tendency for fracture to dominate shear band instability and failure in steel under dynamic torsion as applied strain rate increases. 

First consider failure by shear banding.
Equality $\tau_A^{1/m} = \tau_B^{1/m}$ of \eqref{eq:linmomcons} is integrated to any time $t = t_a > 0$,
with integration limits $\gamma_A = \gamma (Y_A,t_a)$ and $\gamma_B = \gamma(Y_B,t_a)$, initial conditions $\chi_{0A} = \chi_0(Y_A)$, $\chi_{0B} = \chi_0(Y_B)$, and flow stress function \eqref{eq:flowstress}.
Then changing variables and dividing by $g_0^{1/m}/ \dot{\gamma}_0$, 
\begin{align}
\label{eq:Lc1}
& \int_0^{t_a} [g_0 \chi_{0A} \omega(\phi(\gamma(Y_A,t)) \iota^\xi ( \xi(\gamma(Y_A,t)) ]^{1/m}[ 1+ \gamma(Y_A,t)/\gamma_0]^{n/m}[\theta (\gamma(Y_A,t))/\theta_0]^{\nu/m} \dot{\gamma}_0^{-1} \dot{\gamma} (Y_A,t) \d t 
\nonumber
\\ 
&
= \int_0^{t_a} [g_0 \chi_{0B} \omega(\phi(\gamma(Y_B,t)) \iota^\xi ( \xi(\gamma(Y_B,t)) ]^{1/m}[ 1+ \gamma(Y_B,t)/\gamma_0]^{n/m}[\theta (\gamma(Y_B,t))/\theta_0]^{\nu/m} \dot{\gamma}_0^{-1} \dot{\gamma} (Y_B,t) \d t 
\nonumber
\\
&
\Rightarrow 
 \int_0^{\gamma_A}\chi_{0A}^{1/m} \{ \omega(\phi(\gamma)) \iota^\xi(\xi (\gamma))  \}^{1/m}( 1+ \gamma/\gamma_0)^{n/m}[\theta (\gamma)/\theta_0]^{\nu/m}   \d \gamma  
\nonumber
\\ 
&
\qquad \qquad =  \int_0^{\gamma_B}\chi_{0B}^{1/m} \{ \omega(\phi(\gamma)) \iota^\xi(\xi (\gamma))  \}^{1/m}( 1+ \gamma/\gamma_0)^{n/m}[\theta (\gamma)/\theta_0]^{\nu/m}   \d \gamma.  
\end{align}
Dependence of $\xi$ and $\phi$ on $Y$ independently of $\gamma$ is permitted but is not written explicitly in \eqref{eq:Lc1}.
If $L_\infty$ localization occurs, $t_a \rightarrow t_c$, $\gamma_B \rightarrow \infty$, and $\gamma_A \rightarrow \gamma_{Ac}$, where $\gamma_{Ac} > 0$ is finite:
\begin{align}
\label{eq:Lc2}
 & \int_0^{\gamma_{Ac}}\chi_{0A}^{1/m}  \{ \omega(\phi(\gamma)) \iota^\xi(\xi (\gamma))  \}^{1/m}( 1+ \gamma/\gamma_0)^{n/m}[\theta (\gamma)/\theta_0]^{\nu/m}   \d \gamma  
\nonumber
\\ 
&
 =  \underset{\gamma_B \rightarrow \infty}{\lim} \int_0^{\gamma_B}\chi_{0B}^{1/m} \{ \omega(\phi(\gamma)) \iota^\xi(\xi (\gamma))  \}^{1/m}( 1+ \gamma/\gamma_0)^{n/m}[\theta (\gamma)/\theta_0]^{\nu/m}   \d \gamma 
=    \underset{\gamma_B \rightarrow \infty}{\lim} \int_0^{\gamma_B} I(\gamma) \d \gamma.
\end{align}

As explained in Refs.~\cite{molinari1986,molinari1987}, because $\gamma_{Ac}$ is finite by definition, the integrals on the left, and thus the right, sides of \eqref{eq:Lc2} must all be bounded. This means $L_\infty$ localization occurs if and only if $I(\gamma)$ is integrable as $\gamma \rightarrow \infty$.
In Refs.~\cite{molinari1986,molinari1987,claytonJMPS2024}, bounds on viscoplastic properties $n$, $\nu$, and $m$
were derived for which $L_\infty$ localization (i.e., shear band failure) is possible. These prior derivations, based
on power-law viscoplasticity and analytical solutions for $\theta = \theta(\gamma)$, did not consider shear fracture or melting. In the current setting, three possibilities emerge:
\begin{enumerate}  \setlength\itemsep{-0.1em} 
\item Order parameters $\phi$ or $\xi$ evolve with $\gamma$ as $\gamma \rightarrow \infty$.
No definite criteria for the possibility of $L_\infty$ localization are derived since $\theta$, $\omega$,
and $\iota^\xi$ depend on $\gamma$ in forms not known analytically.
\item Either (or both of) $\phi \rightarrow 1$ or $\xi \rightarrow 1$ occurs at finite $\gamma_B$ with $\omega_1 = 0$ or
$\iota^\xi_1 = 0$. In this case, fracture or melt failure precedes shear band failure; the latter never occurs since
$\gamma_B$ is finite.
\item Both $\phi$ and $\xi$ attain fixed terminal values less than unity, or attain unit values with $\omega_1 > 0$ and
$\iota^\xi_1 > 0$, at finite $\gamma_B$. In this case, since $\phi$ and $\xi$ cease to evolve, they do not influence
behavior of $\theta(\gamma)$ in \eqref{eq:thetanomelt} as $\gamma \rightarrow \infty$, and $\omega$ and $\iota^\phi$ become nonzero constants. Thus, the original criterion for the possibility of $L_\infty$ failure applies (see derivations in Refs.~\cite{molinari1986,molinari1987,claytonJMPS2024}):
\begin{align}
\label{eq:Linfty}
  \nu + n + (1-\nu) m < 0, \qquad [ \forall \, m> 0, \nu < 1],
\end{align}
most valid for $m \ll 1$. A Newtonian fluid is recovered for $n = 0$ and $m = 1$, whereby
\eqref{eq:Linfty} is violated. A slightly different analysis of non-hardening ($n = 0$) materials \cite{molinari1987} gives the localization criterion $\nu + m < 0$, also violated for any Newtonian fluid with $\nu \geq -1$. Hence, Newtonian viscosity of molten material must be omitted for $L_\infty$ failure, as done herein.
\end{enumerate}

Next consider failure by fracture in the context of \eqref{eq:phisol2}.
If $\omega_1 > 0$, shear fracture is possible at finite $\gamma$, necessarily preceding
failure by shear banding or melting. However, as some strength is maintained when $\omega_1 > 0$,
residual material can still fail by melting, and possibly, shear banding.
If $\omega_1 = 0$, then $\gamma \rightarrow \infty \Rightarrow W^P_f \rightarrow \infty \Rightarrow \phi \rightarrow 1$. Failure by shear fracture, melting, and potential shear banding can take place concurrently as $\gamma \rightarrow \infty$, but criterion \eqref{eq:Linfty} does not necessarily hold.

Lastly consider failure by melting in the context of heuristic model \eqref{eq:xisol2} with $\gamma_0^\xi = 0$,
$h_T > 0$, 
and $a_\xi = \text{constant}$. Since $\bar{R}$ is non-negative by \eqref{eq:chi},
$\theta(\gamma) \rightarrow \infty \Rightarrow \Xi \rightarrow \infty \Rightarrow \xi \rightarrow 1$.
It is anticipated, but not proven, that melt failure occurs iff $\gamma \rightarrow \infty$ with $\omega > 0$,
meaning complete melting at infinite strain if complete strength loss from fracture has not occurred already.
In contrast, if $\omega \rightarrow 0 \Rightarrow \tau \rightarrow 0$ at finite $\gamma$, then
failure by melting, like failure by shear banding, does not arise, since plastic dissipation supplying
a temperature rise ceases and $\bar{R}$ becomes constant after shear fracture.
If shear fracture does not occur at finite $\gamma$, then simultaneous shear banding,
melt failure, and shear fracture are possible as $\gamma \rightarrow \infty$, but \eqref{eq:Linfty} need not necessarily apply for shear banding.

\subsection{Homogeneous solutions}
Henceforth, the analysis allows non-uniform $\chi_0 = \chi_0(Y)$ and sets $\theta_i = \theta_0 = \text{constant}$.
All other physical properties are constant over $\Omega_0$.
If $\chi_0$ is uniform, $L_\infty$ localization cannot occur: all points $Y$ are indistinguishable so will share the same stress-strain-temperature history.
Failure by shear fracture and melting remain possible. These could occur at finite or infinite $\gamma$.
For homogeneous conditions, the entire slab would fracture or melt simultaneously.
Average strain $\bar{\gamma}$ and defect parameter $\bar{\chi}_0$ in the slab, whose average strain rate is $\dot{\bar{\gamma}}$ since 
$\upsilon_0 = \text{constant}$, with $\hat{y} = (J^E)^{1/3} Y$, are
\begin{align}
\label{eq:gammabar}
\bar{\gamma}(t) = \frac{\upsilon_0 t}{ h_0} = \frac{1}{h_0} \int_0^{h_0} \gamma (Y(\hat{y})) \d \hat{y},
\qquad
\bar{\chi}_0 = 1 - \delta \bar {\chi}_0 = \frac{1}{h_0} \int_0^{h_0} \chi_0 (Y(\hat{y})) \d \hat{y}.
\end{align}
Factor $(J^E)^{1/3}$ accounts for
$h_0$ being the coordinate of the top of the slab {after} application of $p_0$ in Fig.~\ref{fig1} and kinematic ansatz  \eqref{eq:varphi}--\eqref{eq:FP}.
If $p_0 = 0$, then $\hat{y} = Y$. 

For homogeneous conditions, stress from \eqref{eq:flowstress} is, at fixed $p_0$, $\dot{\gamma} = \dot{\bar{\gamma}}$, and $\chi_0 = \bar{\chi}_0$,
\begin{align}
\label{eq:flowstress3}
\bar{\tau}(\bar{\gamma},\xi(\bar{\gamma}),\phi (\bar{\gamma}),\theta (\bar{\gamma})) = g_0 \bar{\chi}_0
\iota^\xi( \xi(\bar{\gamma})) \omega ( \phi(\bar{\gamma}))  (1 + \bar{\gamma} / \gamma_0)^n [\theta(\bar{\gamma}) / \theta_0] ^\nu [\dot{\bar{\gamma}}/ \dot{\gamma}_0]^m. 
\end{align}
Order parameters and temperature are uniform, but functions $\phi(\gamma)$, $\xi(\gamma)$, and $\theta(\gamma)$ are not known analytically except in degenerate cases. Rather, \eqref{eq:phidot2}, \eqref{eq:dxi}, and \eqref{eq:Tdot4} comprise
a set of three coupled ODEs to be integrated numerically, concurrently with \eqref{eq:flowstress3}, over the strain history $\gamma(t)$:
\begin{align}
\label{eq:ODEs}
{ \d \phi} /{ \d \gamma} = {\mathcal F}_\phi (\gamma,\xi,\phi,\theta), \qquad
{ \d \xi}/ { \d \gamma} = {\mathcal F}_\xi (\gamma,\xi,\phi,\theta), \qquad
{ \d \theta} /{ \d \gamma} = {\mathcal F}_\theta (\gamma,\xi,\phi,\theta).
\end{align}
Initial conditions prescribed for \eqref{eq:ODEs} are $\phi(0) = \xi(0) = \bar{R}_0 = 0$ and $\theta(0) = \theta_0$.

\subsection{Localization calculations}
Among all possible positions $Y$ in the slab, localization ensues at the earliest possible $t_c$, at point $Y = Y_B$, for which the rightmost integral in \eqref{eq:Lc2} is a minimum \cite{molinari1986,molinari1987}.  
Shear band failure, if occurring, commences at $Y_B = {\rm argmax} \{ \delta \chi_0 \}$ where $\chi_{0B}$ is a minimum. This follows from the integrand of $\int_0^{t_c} \tau^{1/m} \d t$ in \eqref{eq:Lc1} being non-negative and approaching zero only as $t \rightarrow t_c \Leftrightarrow \gamma_B \rightarrow \infty $ at $B$. 
This threshold integral for localization at $Y_B$ drops as $\chi_{0B}$ decreases because $\chi_0 > 0$ and $m > 0$.
For $m \ll 1$, the localization integral is very sensitive to the
perturbation $\delta \chi_0$ \cite{molinari1986,molinari1987,claytonJMPS2024}.

A shear band approaches a singular surface at $Y_B$ across which displacement has a jump discontinuity as $\gamma \rightarrow \infty$. It is assumed that $\gamma$ and $\dot{\gamma}$ are continuous functions of $Y$ except at singular point(s) $Y_B$ at $t \geq t_c$.
In the shear band, from \eqref{eq:Lc1} and \eqref{eq:Lc2}, $\tau^{1/m} / \dot{\gamma} \rightarrow 0$
at $B$ as $t \rightarrow t_c$. 
At $t < t_c$, since ${\gamma}(Y)$ is
continuous, $\gamma_A$ at at least one location $Y_A$ (that can in principle vary with $t$) must match ${\bar{\gamma}} = \upsilon_0 t / h_0$.  If this $Y_A$ is time-independent, then $\dot{\gamma}_A = \dot{\bar{\gamma}} = \upsilon_0/h_0$ identically. Stress exactly from \eqref{eq:flowstress} in that scenario, otherwise approximated by \eqref{eq:flowstress2} at each $Y_A$, is given by \eqref{eq:flowstress3} for $t < t_c$.
From \eqref{eq:linmomcons}, this value of $\tau$ is equally valid for the entire $Y$-domain for $t < t_c$.
Equations \eqref{eq:ODEs} are likewise valid to the same order of approximation for $t< t_c$,
with estimate
$\chi_0 \approx \bar{\chi}_0$ in ${\mathcal F}_\xi$ and $\mathcal{F}_\theta$ \cite{molinari1987,molinari1988,claytonJMPS2024}, noting $\mathcal{F}_\phi$ does not depend on $\chi_0$.
With these assumptions, $\dot{\gamma}$ affects the solution only through $\tau$, and since $\tau$ is uniform over $\Omega_0$, it follows that $\theta$, $\phi$, and $\xi$ are suitably approximate functions only of $\gamma$ at each point $Y$,
for given loading and initial conditions $\upsilon_0$, $p_0$, and $\bar{\chi}_0$.
 
Computation of average localization strain $\bar{\gamma}_c$ proceeds as in Ref.~\cite{claytonJMPS2024}. Point $Y_B$ is that for which $Y_B = {\rm argmax} \{ \delta \chi_0(Y) \} = {\rm argmin} \{ \chi_0(Y) \} $.
A numerical value of the right integral in \eqref{eq:Lc2} is found as $\gamma_B \rightarrow \infty$.
This integral converges (diverges) if $L_\infty$ localization is possible (impossible).
If converged, the left of \eqref{eq:Lc2} is set to this value at all $Y_A$ where $\chi_{0A} > \chi_{0B}$
and solved for $\gamma_{Ac}(Y)$ at each $Y \neq Y_B$.
Critical strain $\bar{\gamma}_c$ is lastly found by integrating $\gamma = \gamma_{Ac}(Y)$ over $Y$ in \eqref{eq:gammabar}.
Letting $I(\gamma(Y))$ be any integrand in
\eqref{eq:Lc2} and setting $\gamma_{Ac}(Y) = 0$ at $Y = Y_B$ to exclude singularities \cite{molinari1986,molinari1987},
\begin{equation}
\label{eq:gbform1}
\gamma_{Ac} = \underset{\gamma_A \in (0,\infty)}{\text{arg} \, 0} \left[ {\int_0^{\gamma_{A}} I(\gamma(Y_A))\, \d \gamma }- \underset{\gamma_B \rightarrow \infty}{\lim} \int_0^{\gamma_B} I(\gamma (Y_B)) \, \d \gamma \right], \qquad
\bar{\gamma}_c = \frac{1}{h_0} \int_0^{h_0} \gamma_{Ac} (Y(\hat{y})) \, \d \hat{y}.
\end{equation}
Numerical iteration is used to solve the first of \eqref{eq:gbform1},
and numerical integration to solve the second of \eqref{eq:gbform1}.
The localization threshold integral on the right of \eqref{eq:Lc2},
and thus $\bar{\gamma}_c$ and $t_c$, are affected by transients in $\phi$ and $\xi$; how so
 depends on properties and is not obvious from inspection.
As strain in the vicinity of the band accommodates more of the average, $\gamma$ drops elsewhere for the same $\bar{\gamma}$.

Previous calculations \cite{molinari1986,molinari1987,claytonJMPS2024}
did not seek to model degradation and failure within the adiabatic shear band.
Therein, $\bar{\gamma} = \bar{\gamma}_c$ for $t \geq t_c$, and the
drop to zero stress upon localization was modeled abruptly (e.g., $\bar{\tau} = 0$ for $t > t_c$).
A different approach is taken in the present work, similar to Refs.~\cite{dolinski2010,dolinski2015,dolinski2015b}. These studies likewise used plastic work-threshold based damage models for adiabatic shear failure to capture gradual post-localization stress decay as
witnessed in torsion experiments on iron and steels (e.g., \cite{fellows2001b,fellows2001,marchand1988}).
Recall $\bar{\gamma}_c$ of \eqref{eq:gbform1} 
is the average strain in the slab that numerically excludes the contribution of the core of the fully formed shear band at $Y_B$: this core is infinitesimally wide but supports infinite shear strain $\gamma_{Bc} \rightarrow \infty$. 
The latter's contribution, from a discrete zone wherein shear fracture tends to concentrate, is represented by effective jump in shear displacement $\Delta_\phi (t) \geq 0$
that can be formally derived using Gauss's theorem
\cite{claytonJMPS2005,claytonIJSS2003,claytonMECMAT2004}:
\begin{align} 
\label{eq:deltaphi}
\bar{\gamma} (t) = \bar{\gamma}_c + \Delta_\phi (t) / h_0,   \qquad [\forall \, t  \geq  t_c; \, \Delta_\phi(t) = 0 \, \forall \, t \leq t_c].
\end{align}
Post-localization shear displacement jump $\Delta_\phi$ need not be calculated explicitly, but it can
be using $\bar{\gamma} = \upsilon_0 t / h_0 $ and $\bar{\gamma}_c$ from \eqref{eq:gbform1}. 
The average strain $\bar{\gamma}$ in \eqref{eq:deltaphi} is identified with $\bar{\gamma}$ in \eqref{eq:flowstress3} to calculate stress $\bar{\tau}$ for $ t > t_c \leftrightarrow \bar{\gamma} > \bar{\gamma}_c$ and $\bar{\chi}_0 < 1$.
Physical consistency of this post-localization stress calculation requires that $\bar{\gamma}_c$
match the applied strain needed for ductile shear fracture to initiate under macroscopically homogeneous deformation, as in \eqref{eq:flowstress3} of \S4.2.
Only the average stress decay is continuously depicted for the time history $t > t_c$. Transients of local state variable fields $\gamma$, $\theta$, $\phi$, and $\xi$ are not modeled for the entire time history $t > t_c$.
However, the limit analysis does produce contours $(\cdot)_c$ representing
state variables at the final, failed material state as $ \bar{\tau} \rightarrow 0$ and $t \gg t_c$. 

In some demonstrative calculations of \S5 and \S6, ductile fracture is suppressed by setting threshold $W^\phi_0 \rightarrow \infty$.
In those calculations, $\Delta_\phi = 0$ is imposed and local fields remain static for any $t > t_c$.
For melt failure, this is consistent with neglecting viscosity of the liquid. For shear banding without resolution of ductile degradation,
the predicted abrupt post-localization load drop is consistent with prior numerical implementations of the $L_\infty$ criterion \cite{molinari1986,molinari1987,claytonJMPS2024}.

Defect $\tilde{\phi}_0 (Y) = \delta \chi_0(Y)$, with a 
 distribution assigned as in Refs.~\cite{claytonJMPS2024,le2018}, instigates localization at a single point $Y_B$ at the midsection of the slab $\Omega_0$ in Fig.~\ref{fig1}.
 Recall $\hat{y}$ is the coordinate at $t = 0^+$ after application of $p_0$ but preceding shear by $\gamma(t)$.  With $h_0$ identified in Fig.~\ref{fig1}, define $\tilde{y}$ having its coordinate origin at the midpoint of $\Omega_0$: $\hat{y} = \frac{1}{2} h_0 \leftrightarrow  \tilde{y} = 0$. The initial defect distribution (i.e., yield strength perturbation) of intensity $\epsilon_0$ and width $\lambda_0$ is prescribed in dimensionless form as  \cite{claytonJMPS2024,le2018}
  \begin{align}
 \label{eq:deltachi}
  & \delta \chi_0 (\tilde{y}(Y)) = \epsilon_0 \exp {[} - {(2\tilde{y}/\lambda_0)^2} {]},
  \qquad [0 < \lambda_0 \leq 1, \epsilon_0 \geq 0];
 \nonumber
 \\
& \tilde{y}(Y) = ({2}/{h_0}) {[} \hat{y}(Y) - {h_0}/{2} {]} = 
( {2}/{h_0}){[} (J^E(p_0))^{1/3} Y - {h_0}/{2} {]} \in [-1,1] ;
 \nonumber 
 \\
 & \delta \bar{\chi}_0 = 1- \bar{\chi}_0 =  \frac{1}{2}\int_{-1}^1 \delta \chi_0 \, \d \tilde{y}
 = \frac {\sqrt{\pi}}{4} \epsilon_0 \lambda_0 \, {\rm erf} \biggr{[} \frac{2}{\lambda_0} \biggr{]} \approx 
 0.4431 \, \epsilon_0 \lambda_0 .
 \end{align}
 In calculations, $\gamma$ and $y$ domains are discretized into dimensionless increments  $\d {\gamma}$ and $\d \tilde{y}$. Taking $\d \gamma, \d \tilde{y} \lesssim 10^{-4}$ is generally sufficient 
 for $\bar{\gamma}_c$ independent of grid size.
With numerical integration limited by machine precision and an infinite number of increments $\d \gamma$ impossible, a large enough upper bound on $\gamma_B$ is consistently used
 in \eqref{eq:gbform1} to enable
 respectable convergence of $\int_0^{\gamma_B} I (\gamma) \d \gamma$ toward a constant value if enabled by properties (e.g., \eqref{eq:Linfty}) and loading conditions.
 Convergence quickens with fracture or melting: $(\iota^\xi \omega)^{1/m} \rightarrow 0$
 rapidly with $\gamma$ as $\phi(\gamma) \rightarrow 1$ or $\xi(\gamma) \rightarrow 1$  for $m \ll 1$. In such cases, if failure by shear fracture or melting occurs, \eqref{eq:gbform1} is still used
 with singular point $Y_B$ excluded from $\bar{\gamma}_c$.  Infinite strain remains
 possible at $Y_B$ if material has null strength there.
 
 Numerical results include contours of $\gamma_c$, $\theta_c$, $\phi_c$, and $\xi_c$, where subscript $(\cdot)_c$
denotes a quantity as $t \gg t_c$ and $\bar{\tau} \rightarrow 0$. Contours are restricted to dimensionless $\tilde{y}$-space. They enable comparison of {dimensionless} widths of high-strain, high-temperature
zones centered on singular bands at $\tilde{y} = 0$. 
Absolute zone widths scale linearly with $h_0$ if $t_0 = h_0/ \upsilon_0$, $\epsilon_0$, and $\lambda_0$ are all constants.
Minimum widths tend to zero as $h_0 \rightarrow 0$ at fixed $\lambda_0$ since gradient regularization, conduction, and inertia are not modeled.
To account for regularizing physics and predict absolute shear band widths, more sophisticated numerical methods (e.g., finite difference \cite{cherukuri1995} or finite element \cite{mcauliffe2016}) are required.

\section{Application to steel \label{sec5}}

\subsection{Properties and parameters}
Behavior of a high-strength, low C, Ni-Cr steel (i.e., a type of RHA steel)
is analyzed.
Mechanical properties follow from experimental studies of the 1970s-1990s~\cite{benck1976,hauver1976,hauver1979,moss1981,gray1994},
and for viscoplastic response, model calibrations to experimental data \cite{meyer2001,claytonJMPS2024}.
A typical composition \cite{benck1976,moss1981} has 0.22 wt.\% C, 1.06 wt.\% Cr, and 3.15 wt.\% Ni, with other trace elements.
 Material properties and parameters are listed in Table~\ref{table1}.
 Most properties, and most viscoplastic parameters,
 are repeated from Ref.~\cite{claytonJMPS2024} where original
  sources and calibrations can be found.
Exceptions are static yield strength $g_0$ and rate sensitivity $m$.
Values in Refs.~\cite{meyer2001,claytonJMPS2024} provided a best-fit to dynamic compression
data optimized for strain rates of $10^4$/s. Experiments modeled in the current work \cite{fellows2001b} consider a Ni-Cr
steel of RHA type, but at lower average rates (e.g., $\dot{\bar{\gamma}} \approx 3200$/s) and for which a lower flow stress was observed
than is predicted via past values $g_0 = 0.693$ GPa and $m= 0.065$. 
Iron and steels show increasing $m$ for strain rates exceeding around 10$^3$/s due to
influences of phonon drag and relativistic mechanisms at very high rates \cite{kocks1975,claytonNCM2011,claytonJDBM2021,sadjadpour2015,gur2018}.
Logically, $m$ is reduced to better represent rates $\lesssim 10^4$/s; then, $g_0$
is calibrated to match the experimental torsion data \cite{fellows2001b} for $\bar{\gamma} < \bar{\gamma}_c$.

Like pure Fe, the $\alpha$, $\epsilon$, and $\gamma$ phases of this steel are BCC, HCP, and FCC.
At $p = 0$ and $\theta_0 = 300$ K, material is fully of $\alpha$ phase.
Near $\theta_0$, $\alpha \rightarrow \epsilon$ transformation occurs at $p \approx 13$ GPa,
and near zero pressure, $\alpha \rightarrow \gamma$ transformation occurs at $\theta \approx 1000$ K
(see Ref.~\cite{claytonJMPS2024} and works cited therein).
The present treatment is restricted to $p_0 < 13$ GPa, but shear-induced $\alpha \rightarrow \epsilon$ transformations, though not reported in known experiments on RHA, might be possible
since they have been inferred for Fe \cite{rittel2006}.
Evidence of the $\alpha \rightarrow \gamma$ transition within shear bands has been
noted in some experiments \cite{cho1990,syn2005}, but not others \cite{moss1981,fellows2001b}. To keep the
analysis and interpretation of results tractable, $\alpha \rightarrow \epsilon/ \gamma$ transitions are not modeled herein.
If solid-solid transformations do occur, the analysis amounts to assigning each solid phase the same
 properties and ignoring any dissipation from transformations.
 
  For localization calculations, $\lambda_0$ follows from Ref.~\cite{claytonJMPS2024}.
 Effects of the initial defect width in \eqref{eq:deltachi} are explored in \S5.2 by two choices $\lambda_0 = 0.5$ and $\lambda_0 =1.0$. 
 In Refs.~\cite{molinari1986,molinari1987,claytonJMPS2024}, ranges of perturbation intensity spanning
 $\epsilon_0 \in [10^{-6},10^{-1}]$ were considered in parametric calculations on steels and Fe.
 Therein, $\bar{\gamma}_c$ decreased approximately linearly with $\log_{10} \epsilon_0$.
 In Table~\ref{table1}, $\epsilon_0 = 10^{-3}$, intermediate among prior cited values,
 is suitable in \S5.2 for matching $\bar{\gamma}_c$ to average shear at localization from experiment \cite{fellows2001b} when $\lambda_0 = 0.5$. 
 However, as also shown in \S5.2, $\epsilon_0$ is not always a free
 parameter: $\epsilon_0$, $\lambda_0$, $W^\phi_0$, and $\bar{\gamma}_c$ are not all independent.  
 In \eqref{eq:Linfty}, $ \nu + n + (1-\nu) m = -0.233 < 0$, so
 shear band failure is possible for this steel, at least in the absence of melting or fracture.
  
 New parameters not repeated from Ref.~\cite{claytonJMPS2024}
 are those for the shear fracture model of \S3.4 and the melting model of \S3.5.
 For dynamic shear fracture, the simplest physically admissible choices entering
  \eqref{eq:Ren}, \eqref{eq:omega}, and parameters consistent with experimental
 data \cite{fellows2001b} modeled in \S5.2 are assumed.
 These include $f_\phi = \phi$, $\omega_1 = 0$, $k = 1$, $\iota^\phi = 1 - \phi$, and $\gamma^\phi_0 = 0$
 as in example \eqref{eq:phisol2}.
 The latter assumes isotropic ductile damage (e.g., spherical voids);
 detailed microscopy data would be needed to justify a nonzero value.
 The presently consulted data \cite{fellows2001b} do not allow independent 
 calibration of $\gamma^\phi_0$ and $E_C$; setting the former to zero is the
 simplest choice.
 The value $W^{\phi}_0 =  450$ MPa is taken verbatim
 from Ref.~\cite{dolinski2015} on RHA steel, with the caveat that
 values ranging from 20 to 350 MPa have been used elsewhere for
 various structural steels \cite{wang2020,zeng2022}.
 Phase-field studies \cite{mcauliffe2015,mcauliffe2016,wang2020,zeng2022}
 consistently assign 10\% of the plastic work
 to energy consumed by fracture, giving $\alpha_0 \beta_0 = 0.1$ in Table~\ref{table1}.
 Cohesive energy $E_C$ is unknown a priori for the specific steel of present study;
 the value $E_C = 25$ MPa in Table~\ref{table1} is calibrated in \S5.2 to the post-localization stress-strain data of Ref.~\cite{fellows2001b}.
 Values ranging anywhere from 10 to 1600 MPa can be inferred
 from experimental and phase-field studies on other high-strength
 steels \cite{minnaar1998,mcauliffe2015,mcauliffe2016,wang2020,zeng2022}.
 Induced porosity at fracture quantified by $\delta^\phi_0$ is 
 difficult to quantify precisely for dynamic shear, but voids and cracks have been observed inside shear bands
 in experiments on steels \cite{cho1990,minnaar1998,fellows2001b}.
 A representative value of 0.05 from Ref.~\cite{claytonZAMM2024} on Fe and other steels is used here.
 With this choice and the value of $E_C$ in Table~\ref{table1}, 
 tensile pressure that would induce rupture is $-p_0 \approx E_C / \delta_0^\xi = 0.5 $ GPa.
 The corresponding uniaxial tensile stress would be $\approx 1.5$ GPa.
 For reference, the static ultimate tensile strength of RHA steel ranges from 0.8 to 1.2 GPa \cite{benck1976,singh2021},
 and spall strengths of Fe and steels range from 0.7 to 3.7 GPa \cite{wang2017}. 
  
 For melting, $\xi \rightarrow \bar{\xi}$, $\gamma^\xi_0 = 0$, and \eqref{eq:interpxi} is used.
 Metastable equilibrium, $t_R^\xi  \dot{\bar{\gamma}} \ll 1$,
 is justified by suitability of shock-wave experiments that complete at timescales on the
 order of $\mu$s to determine melting curves of Fe \cite{belo2000,nguyen2004}.
 Omission of $\gamma^\xi_0$ is the usual assumption \cite{hwang2014,hwang2016}, satisfactory in the limit of sharp interfaces \cite{levitas2011b,grinfeld1991}.
 A nonzero value could be important for modeling nanoscale physics
 in a regularized phase-field theory \cite{levitas2011b}, but calibration to nm-scale data from experiments or atomic simulations is needed.
   The equilibrium melt temperature at ambient pressure, $\theta_T$, is around 1800 K
 in Fe \cite{tan1990} and the present class of steels \cite{gray1994,meyer2001}.
Unknown for the current Ni-Cr steel, 
 values of $h_T$ and $\delta^\xi_0$ for Fe \cite{tan1990} are used
 as a substitute.
 The model can be considered reasonably accurate only for $p_0 \lesssim10$ GPa since a near-linear dependence of melt start temperature
 on pressure is predicted by \eqref{eq:interpxi}.
 At higher $p_0$, experiments and atomic simulation data
 show highly nonlinear melt curves for Fe  \cite{belo2000,nguyen2004}.
 Such data also show mixed-phase regions where liquid and solid coexist
 at a given $\theta$ and $p_0$.
 The width of the mixed-phase region, and dissipation from melting, are controlled
 by $\beta_\xi$, for which precise values are unknown. 
 Assumed in Table~\ref{table1} are two choices for $\beta_\xi$: one, for ``fast melting'' with 
 $\beta_\xi = 0.18$ GPa, one for ``slow melting'' with $\beta_\xi = 1.21$ GPa.
 For the former, under purely thermal loading, $\xi = 0.99$ at $\frac{\theta}{\theta_T} = 1.2$.
 For the latter, $\xi = 0.5$ at $\frac{\theta}{\theta_T}= 1.2$.
 Redundant in the present setting, null values $a_\xi = 0$ and $\iota^\xi_1 = 0$ are assigned.
 Admissible minimum and maximum values of $r_0^\xi$ are explored in calculations: $r_0^\xi = 0$ and $r_0^\xi = 1$.
   
 \begin{table}
\footnotesize
\caption{Properties or model parameters for Ni-Cr steel, $\theta_0 = 300$ K}
\label{table1}       
\centering
\begin{tabular}{lll | lll}
\hline\noalign{\smallskip}
Parameter [units] & Definition &Value $\quad $ & $\quad$ Parameter [units] & Definition &Value  \\
\noalign{\smallskip}\hline\noalign{\smallskip}
$\rho_0$ [g/cm$^3$] & ambient mass density &  7.84   & $\quad$ $c_V$ [MPa/K] & specific heat & 3.48  \\  
${B}_0$ [GPa] & isothermal bulk modulus   & 163      & $\quad$ $h_T$ [GPa] & latent heat of fusion  &  2.09  \\   
${B}'_0$ [-] & pressure derivative of $B$  & 5.29    & $\quad$ $\delta^\xi_0$ [-] & melt volume change  & 0.052 \\ 
$\mu_0$ [GPa] & elastic shear modulus & 80  & $\quad$ $ \theta_T$ [K]  & equilibrium melt & 1800 \\
$\mu'/\mu_0$ [1/GPa] & pressure dependence of $\mu$  & 0.024 & $\quad$ $ \beta_\xi$ [GPa] & melt kinetic barrier &  0.18, 1.21 \\   
$ g_0$ [GPa] & initial yield strength & 0.40 & $\quad$ $W^{\phi}_0$ [GPa] & damage start threshold & 0.45 \\ 
$ \gamma_0$ [-] & reference plastic strain  & 0.01 & $\quad$ $ \alpha_0$ [-] & fracture energy ratio & 0.125  \\   
$\dot{ \gamma}_0$ [1/s] & reference strain rate  & 1.0 & $\quad$ $E_C$ [GPa] & fracture cohesive energy & 0.025 \\ 
 $n$ [-] & strain hardening exponent  & 0.05  & $\quad$ $\delta_0^\phi $ [-] & porosity at failure & 0.05 \\
 $m$ [-] & strain rate sensitivity  & 0.035  & $\quad$ $\omega_1$ [-] & residual strength factor & 10$^{-5}$ \\  
 $\nu$ [-] & thermal softening & -0.33   & $\quad$ $\epsilon_0$ [-] & nominal defect intensity & 10$^{-3}$ \\   
  $ {\beta}_0$ [-] & Taylor-Quinney factor & 0.8   & $\quad$  $\lambda_0$ [-] & initial defect width & 0.5, 1.0 \\   
\noalign{\smallskip}\hline
\end{tabular}
\end{table}

 \subsection{Numerical results}
 
 Outcomes of calculations invoking methods of \S4.2 and \S4.3, with baseline properties of Ni-Cr steel in
 Table~\ref{table1}, are reported next.  Results in \S5.2 consider only null external pressure (i.e., $p_0 = 0 \Leftrightarrow J^E = 1$).
 Results are compared with dynamic torsion data of Ref.~\cite{fellows2001b}, specifically data from the
 experiment on specimen labeled ``B6''.  In that work~\cite{fellows2001b}, a torsional split Hopkinson
 (i.e., Kolsky) bar was used to determine average dynamic shear stress versus shear strain behavior,
 and local strains and strain rates were calculated by analyzing high-speed photographs taken in-situ on
 portions of specimens marked beforehand with regularly spaced grids.
 Data for six specimens of 2-mm length were reported, but local strain distributions were only shown for B6 (hence the reason
 for modeling that particular sample).
 Among the six samples, average wall thickness varied by $\approx 10$\%, mean strain rates varied up to a factor of two, and localization was observed
 in four of the six experiments.
 Among those four, estimated times and strains at localization onset varied by around 50\%.
 As the material is not extremely rate sensitive, test-to-test variations are attributed to different
 initial defects that could possibly be linked to thickness perturbations \cite{molinari1986,molinari1987,claytonJMPS2024}.
 
 Recall the shearing rate of the present framework is $\dot{\gamma} = 2 \dot{\epsilon}$, where
 in the absence of dilatation, the true local strain rate is $\dot{\epsilon} = \dot{\epsilon}_{xy} = \frac{1}{2}(\partial \upsilon_x / \partial y + \partial \upsilon_y / \partial x)$.  The mean shearing rate
 \footnote{The strain measure in Ref.~\cite{fellows2001b} does not appear to be formally defined therein; here, it is
interpreted as true strain. If, instead, it is nominal strain (with rotation), then experimental values of $\gamma$
 and $\bar{\gamma}$ quoted herein should be multiplied by $\frac{1}{2}$. Threshold energy $W^\phi_0$ would be similarly reduced. Forthcoming trends in model predictions should be unchanged, but adiabatic temperature rise would be smaller. The present true-strain interpretation is supported by consistency of the current model and parameters (i.e., Table~\ref{table1}) with the identical value of $W^\phi_0$ quoted for RHA in Ref.~\cite{dolinski2015}. }
  for test B6 is taken as $\dot{\bar{\gamma}} = 3200$/s, depicting
 the average reported by the Hopkinson bar analysis \cite{fellows2001b}.
 After localization begins, the mean shear strain rate remains reasonably constant according to the Hopkinson bar analysis,
 but image analysis suggests it could vary \cite{fellows2001b}.  The former experimental result is used herein;
 the framework in \S4 requires $\dot{\bar{\gamma}} = \text{constant}$, and further omits inertial effects
 that would seem more prominent if the average, in addition to local, strain rate fluctuated strongly in time.
 Total average strain from experiment B6 is converted from a start time $t = 0$ when the experimental load-time
 history provided by the Hopkinson bar begins. During the early part of the test (i.e., first $\approx 20 \, \mu$s \cite{fellows2001b}),
 the stress response is overly compliant and highly oscillatory because the sample has neither loaded fully nor achieved
 equilibrium needed for a (nearly) uniform stress-strain state \cite{gray2000}.
 Thus, the oscillatory quasi-elastic offset ``strain'' $\gamma_o = 3200 \times 20 \times 10^{-6} = 0.064$ contributes to the total calculated this way.  For comparison of average stress-strain data with results of the rigid-viscoplastic framework of \S3 and \S4,
 this offset must equivalently be subtracted from the experimental record or added to the model reference datum, since prior to 20 $\mu$s, the bulk of the sample has not plastically yielded. Early-time and fine-scale oscillations are omitted \cite{dolinski2015b}.
    
 \begin{figure}
\begin{center}
 \subfigure[average stress vs.~strain]{\includegraphics[width=0.32\textwidth]{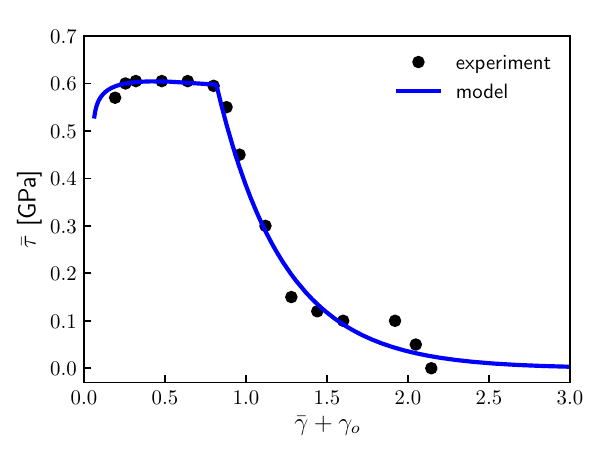} \label{fig2a}} 
 \subfigure[localized strain profile]{\includegraphics[width=0.32\textwidth]{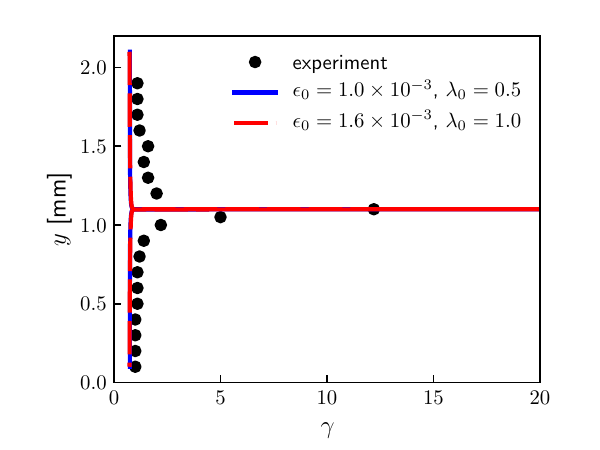}\label{fig2b}} 
  \subfigure[initial defects]{\includegraphics[width=0.32\textwidth]{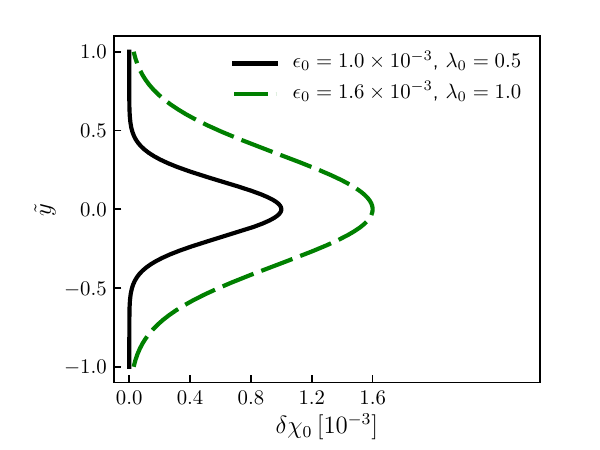}\label{fig2c}} 
  \end{center}
  \vspace{-0.5cm}
\caption{Model results and experimental torsion data \cite{fellows2001b}:
(a) average shear stress $\bar{\tau}$ vs.~average shear strain $\bar{\gamma}$ (strain in rigid-plastic model shifted by $\gamma_o$ to compensate for oscillatory elastic ramp-up in experiments, not shown)
(b) localized strain distribution $\gamma$, specimen size $h_0 = 2$ mm (c) two initial defect profiles ($\epsilon_0,\lambda_0$). Experimental points sample continuous (a) or discrete (b) data from Ref.~\cite{fellows2001b}.}
\label{fig2}       
\end{figure}

Average stress-strain behaviors are compared in Fig.~\ref{fig2a}.
The model result is offset by $\gamma_o$ as it necessarily excludes the oscillatory elastic ramp-up.
The rapid load drop corresponding to the onset of localization appears at $\bar{\gamma} = \bar{\gamma}_c = 0.753$.
Recall from \S5.1 that the model, with prescribed $\epsilon_0 = 10^{-3}$,
is able to accurately match the experimental stress-strain record with calibration of only two
parameters: (i) initial yield strength $g_0$ for average stress up to localization and (ii) cohesive energy $E_C$
for the rate of post-localization stress decay.
As will be shown in parametric calculations of \S6, the stress drop becomes more abrupt as $E_C$ is reduced
or as tensile pressure is increased. 
Localized strain distributions from model ($\gamma = \gamma_c$) and experiment ($ t = 613 \, \mu$s) are compared in Fig.~\ref{fig2b}.
Model results are shifted upward by $\Delta y = 0.1$ mm so that locations of the centers of the shear bands
coincide. This is permissible since $\gamma$-contours are nearly independent of $\tilde{y}$ for $|\tilde{y}| \gtrsim 0.9$.
Local strains are similarly uniform far from the core at $y = 1.1$ mm, but
the model does not capture the strain bulge surrounding the core of the band seen in the experiment.
Post-localization strain within the modeled band is depicted by a finite displacement jump $\Delta_\phi$
via \eqref{eq:deltaphi} over a zero-width (i.e., singular) surface. In the experimental record, the band is of finite width, but cracks not evident in the data profile
were seen around at least some of the perimeter \cite{fellows2001b}.
Discrepancies can be attributed to missing regularization mechanism(s) in the
model (e.g., no conduction, inertia, or gradient surface energies) and
limits of experimental resolution emphasized in Ref.~\cite{fellows2001b}.

The mean stress-strain behavior is the same when a less concentrated
initial defect profile is prescribed, but of higher peak intensity, as depicted in Fig.~\ref{fig2c}.  The value of
$\epsilon_0 = 1.6 \times 10^{-3}$ must be tuned in this case to produce the same
value of $\bar{\gamma}_c = 0.753$: a less concentrated defect requires a larger magnitude to
induce the same average localization strain.

\begin{figure}
\begin{center}
 \subfigure[strain (truncated at $\gamma_c = 1$)]{\includegraphics[width=0.32\textwidth]{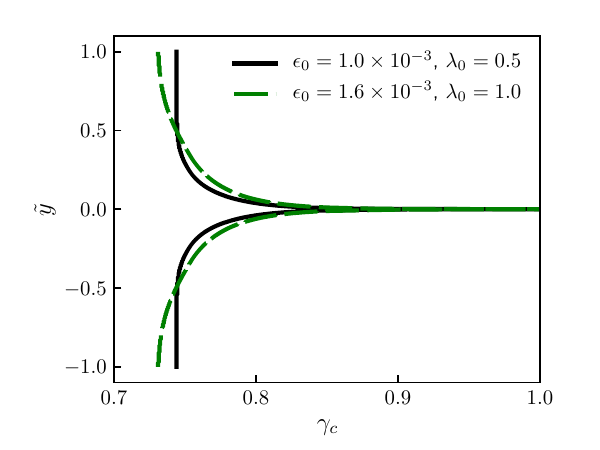} \label{fig3a}} 
 \subfigure[temperature]{\includegraphics[width=0.32\textwidth]{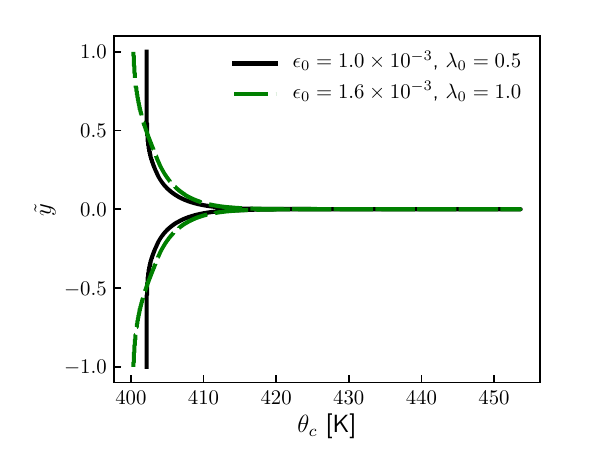}\label{fig3b}}
  \subfigure[shear fracture]{\includegraphics[width=0.32\textwidth]{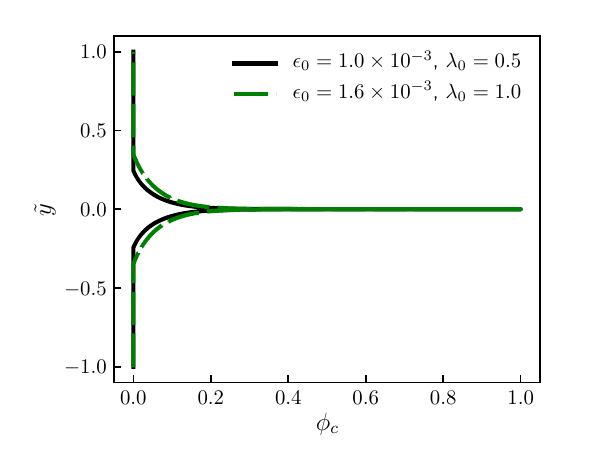}\label{fig3c}} 
  \end{center}
  \vspace{-0.5cm}
\caption{Model results for two initial defect profiles ($\epsilon_0,\lambda_0$) vs.~normalized spatial position $\tilde{y}$:
(a) post-localization strain $\gamma_c$
(b) temperature $\theta_c$ (c) fracture order parameter $\phi_c$}
\label{fig3}       
\end{figure}

Effects of the two defect profiles on critical strain, temperature,
and ductile fracture parameter distributions are further examined in Figs.~\ref{fig3a}, \ref{fig3b},
and \ref{fig3c}.
As anticipated, a more diffuse defect produces a wider strain, temperature, and damage distribution
away from the infinite-strain band centered at $\tilde{y} = 0$ than a narrower initial defect.
Maximum temperature $\theta_c$ does not reach 500 K in either case. Damage-softening
from fracture in the shear band reduces $\tau$ to such a low value that
plastic working ceases to affect the $\theta$ by any pragmatically calculable amount as $\gamma \rightarrow \infty$.
Calculations verify temperature rise is insufficient to induce melting, even at the core of the shear band,
recalling $\theta_T = 1800$ K.
It also appears insufficient to induce $\alpha \rightarrow \gamma$ phase transformation (not enabled in the framework of \S2--\S4) that occurs
at $\approx 1000$ K \cite{moss1981,claytonJMPS2024}.
In the band center, $\phi_c \rightarrow 1$.

\begin{figure}[ht!]
\begin{center}
 \subfigure[shear stress]{\includegraphics[width=0.32\textwidth]{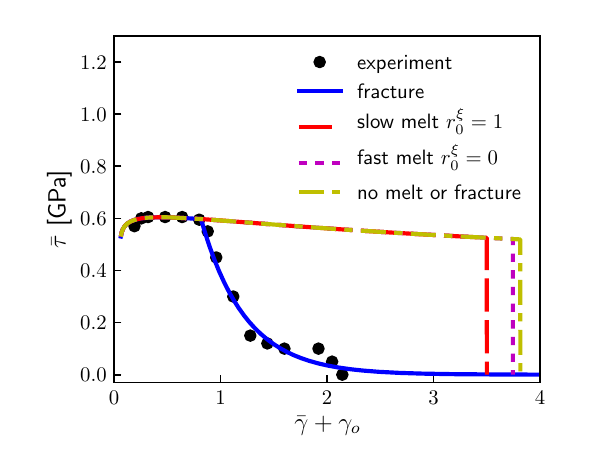} \label{fig4a}} \qquad
 \subfigure[shear stress]{\includegraphics[width=0.32\textwidth]{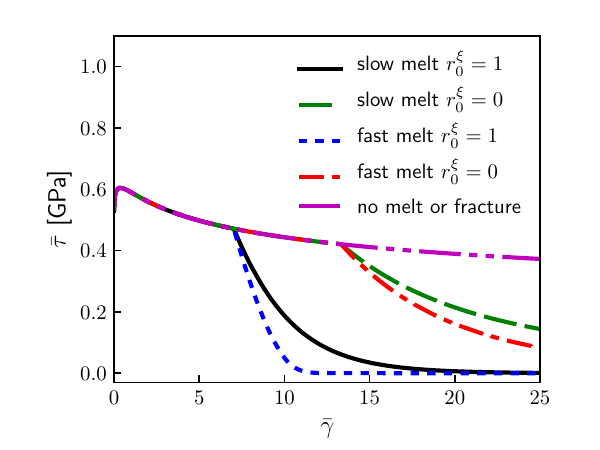}\label{fig4b}} \\
  \subfigure[temperature]{\includegraphics[width=0.32\textwidth]{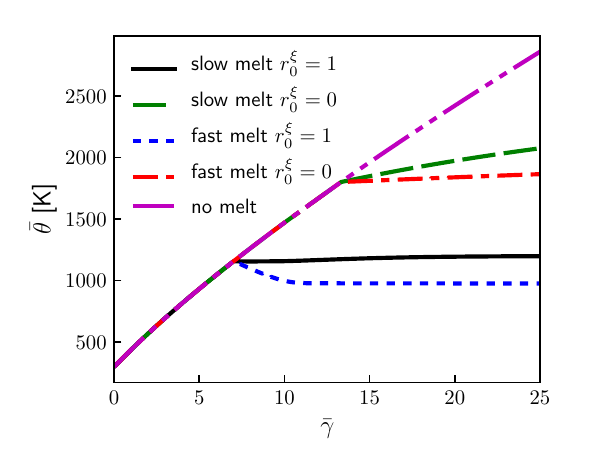}\label{fig4c}} \qquad
    \subfigure[melt fraction]{\includegraphics[width=0.32\textwidth]{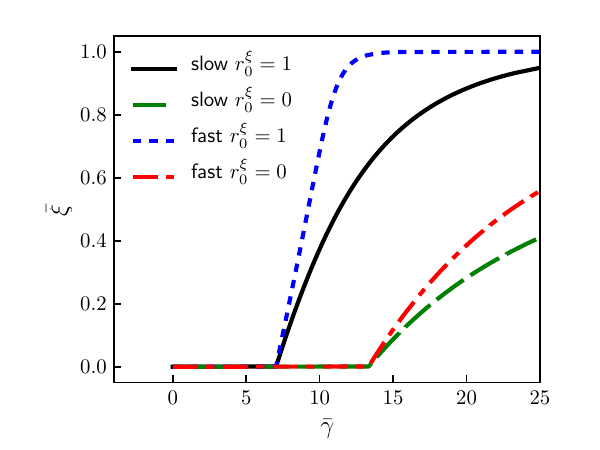}\label{fig4d}} 
  \end{center}
  \vspace{-0.5cm}
\caption{Model results with features activated or suppressed:
(a) average shear stress $\bar{\tau}$ vs.~average shear strain $\bar{\gamma}$
and experimental torsion data \cite{fellows2001b};
 (b), (c), (d) homogeneous solutions for average shear stress,
 temperature $\bar{\theta}$, and melt order parameter $\bar{\xi}$ to extreme shear strain.
 Slow and fast melt correspond to respective large and small values of $\beta_\xi$ in Table~\ref{table1};
 $r_0^\xi$ is fraction of stored energy of cold work released upon melting.}
\label{fig4}       
\end{figure}

Calculations from the full model, fracture suppressed, and
both fracture and melting suppressed are considered in Fig.~\ref{fig4}.
In the full model, labeled ``fracture'' in Fig.~\ref{fig4a}, melting never occurs; results here
are repeated with experimental data from Fig.~\ref{fig2a}.
Cases in which fracture is deactivated but melting enabled are denoted ``fast melt'' and ``slow melt'' for respective
 small and large values of kinetic factor $\beta_\xi$ in Table~\ref{table1}. Different choices of $\beta_\xi$ and $r^\xi_0$ produce modest differences in the total strain at which localization ensues, beyond which the drop in average stress $\bar{\tau}$ is
abrupt per assumptions in \S4.3. The largest critical localization strain is predicted when both softening mechanisms (i.e., fracture and melting) are suppressed.
Figures~\ref{fig4b}, \ref{fig4c}, and \ref{fig4d} compare predictions of model renditions under homogeneous deformations
(i.e., no initial defects: $\epsilon_0 = 0 \leftrightarrow \delta \chi_0 = 0$).
When stored energy of cold work fuels melting, then melting begins at $\bar{\gamma} \approx 7.0$.
When it does not, melting begins at $\bar{\gamma} \approx 13.3$.
Subsequently, melting occurs at a faster rate with respect to $\bar{\gamma}$ for the lower value of $\beta_\xi$,
as evidenced by the homogeneous melt fraction $\bar{\xi}$ in Fig~\ref{fig4d}.
Shear stress decays with increasing $\bar{\xi}$ in Fig.~\ref{fig4b}.
Homogeneous temperature $\bar{\theta}$ increases monotonically when melting is suppressed.
Since melting consumes energy ($h_T > 0$), the rate of temperature increase is reduced during solid-to-liquid transformation. As seen in Fig.~\ref{fig4c}, for the case of fast melting fueled by energy from dissolution of dislocations, $\bar{\theta}$
can even decrease.

\begin{figure}
\begin{center}
 \subfigure[shear strain]{\includegraphics[width=0.32\textwidth]{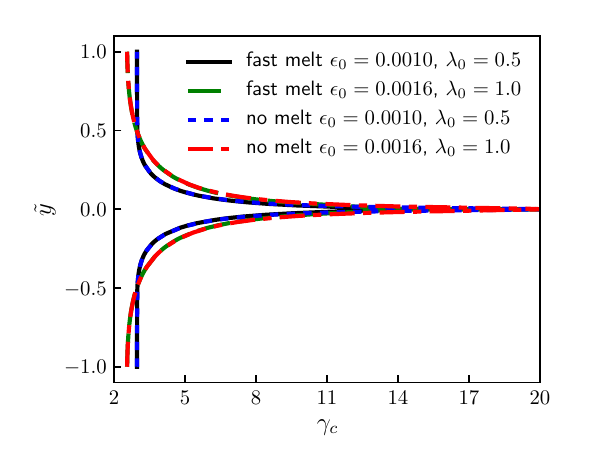} \label{fig5a}} 
 \subfigure[temperature]{\includegraphics[width=0.32\textwidth]{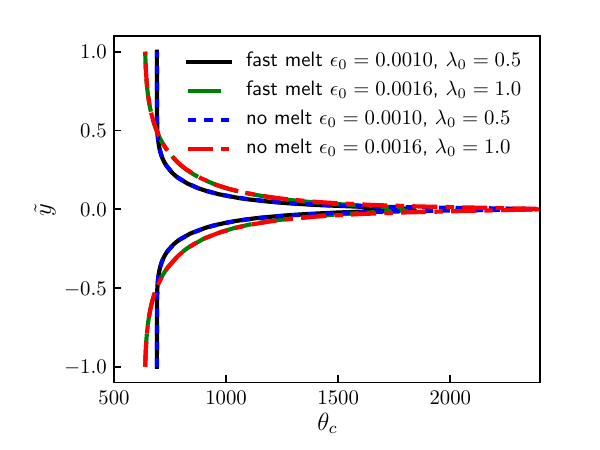}\label{fig5b}}
  \subfigure[melt fraction]{\includegraphics[width=0.32\textwidth]{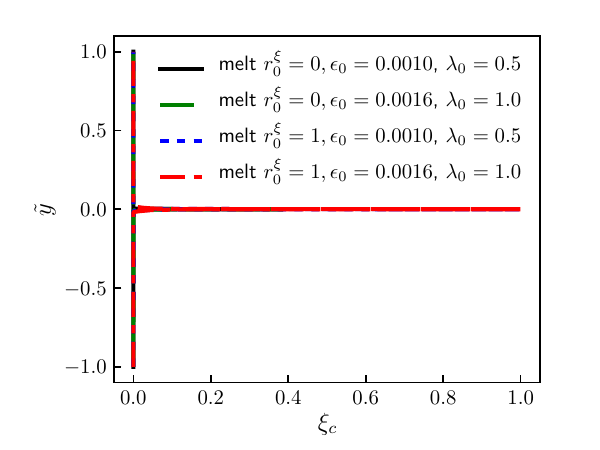}\label{fig5c}} 
  \end{center}
  \vspace{-0.5cm}
\caption{Model results without fracture for two initial defect profiles ($\epsilon_0,\lambda_0$) vs.~normalized spatial position $\tilde{y}$:
(a), (b) strain $\gamma_c$,  temperature $\theta_c$ for $r^\xi_0 = 0$ or melt suppressed
(c) melt parameter $\xi_c$ for $r^\xi_0 = 0,1$}
\label{fig5}       
\end{figure}

Contours of localized shear strain, temperature, and melt fraction are compared in Fig.~\ref{fig5} for different renditions with
fracture suppressed. These correspond to any post-localization time $t > t_c$.
In Fig.~\ref{fig5a}, local strain $\gamma_c$ differs very little whether or not melting occurs. Differences
are larger among the two different choices of $\lambda_0$.
Similar trends are observed for local temperature $\theta_c$ in Fig.~\ref{fig5b}, noting that in the core of the shear band
at $\tilde{y} = 0$, $\theta_c$ can exceed 2400 K.
Accordingly, as seen in Fig.~\ref{fig5c}, complete melting $\xi_c \rightarrow 1$ is possible in the core.
However, temperature and other driving forces (e.g., $\bar{R}$) are not large enough outside the core to 
induce appreciable melting.  Even partially liquified material is always confined to a very narrow region near the center of the band.

Outcomes of calculations of \S5.2 are summarized in Table~\ref{table2}, specifically the average stress $\bar{\tau}_c$
computed at $t = t_c^-$ (i.e., immediately preceding the load drop associated with localization) and the critical average
strain from \eqref{eq:gbform1} 
as $t \rightarrow t_c$ that, by definition, excludes the contribution $\Delta_\phi$ of \eqref{eq:deltaphi}.
Since the steel is already thermal softening prior to any localization, lower $\bar{\gamma}_c$ always correlates with larger $\bar{\tau}_c$.
A more diffuse defect with larger $\lambda_0$ tends to delay localization.
For the same initial defect profile, critical localization strain is reduced 80\% by ductile shear fracture, but by no more than 12\% by melting.
Melting never occurs when shear fracture is permitted.

\begin{table}[ht!]
\footnotesize
\caption{Results summary for high-strength steel, $p_0 = 0$, $\theta_0 = 300$ K, $\dot{\bar{\gamma}} = 3200$/s}
\label{table2}       
\centering
\begin{tabular}{lcccccccc}
\hline\noalign{\smallskip}
Physics		& $W^\phi_0$ [MPa] & $\theta_T$ [K] & $\epsilon_0$ [$10^{-3}$] & $\lambda_0$ & $\beta_\xi$ [GPa] & $r^\xi_0$ & $\bar{\tau}_c$ [MPa] &  $\bar{\gamma}_c$ \\
\noalign{\smallskip}\hline\noalign{\smallskip}
Fracture	& 450		&	1800	&	 1.0		&	 0.5	   &   0.182	       &	1 & 	597				& 0.753	\\	
(full model)	&  		&		&	 1.6		&	 1.0	   &   0.182	       &	1 &	596				& 0.753\vspace{2mm} \\
Melting		& $\infty$	&	1800	&        1.0		&	 0.5	   &   0.182	       &	1 &	526				& 3.432 \\
(no fracture)   & 		&		&        1.6		&	 1.0	   &   0.182	       &	1 &	524				& 3.515 \\
   		& 		&		&        1.0		&	 0.5	   &   1.207	       &	1 &	525				& 3.437 \\
   		& 		&		&        1.6		&	 1.0	   &   1.207	       &	1 &	524				& 3.521 \\
		& 		&		&        1.0		&	 0.5	   &   0.182	       &	0 &	521				& 3.681 \\
		& 		&		&        1.6		&	 1.0	   &   0.182	       &	0 &	517				& 3.874 \\
   		& 		&		&        1.0		&	 0.5	   &   1.207	       &	0 &	521				& 3.683 \\
   		& 		&		&        1.6		&	 1.0	   &   1.207	       &	0 &	517				& 3.877\vspace{2mm} \\
No fracture 	& $\infty$	&   $\infty$	&	 1.0		&	 0.5	   &   0.182	       &	1 & 	519				& 3.751	\\	
or melting		&  		&		&	 1.6		&	 1.0	   &   0.182	       &	1 &	515				& 3.983 \\
\noalign{\smallskip}\hline
\end{tabular}
\end{table}

\section{Parameter variations \label{sec6}}

Parametric calculations next investigate influences of pressure $p_0$ and
values of $W^\phi_0$ and $E_C$ entering the ductile fracture theory of \S3.4.
In all calculations of \S6, the defect width $\lambda_0 = 0.5$ per Table~\ref{table1},
but the intensity $\epsilon_0$ can depart from its nominal value of $10^{-3}$ in Table~\ref{table1}.
Quantities labeled by an  asterisk $(\cdot)^*$ correspond
to nominal property values from Table~\ref{table1} or to results obtained from calculations
using such nominal property values.  Pressures are always limited to $|p_0| \leq 5$ GPa.

\begin{figure}
\begin{center}
 \subfigure[critical localization strain]{\includegraphics[width=0.32\textwidth]{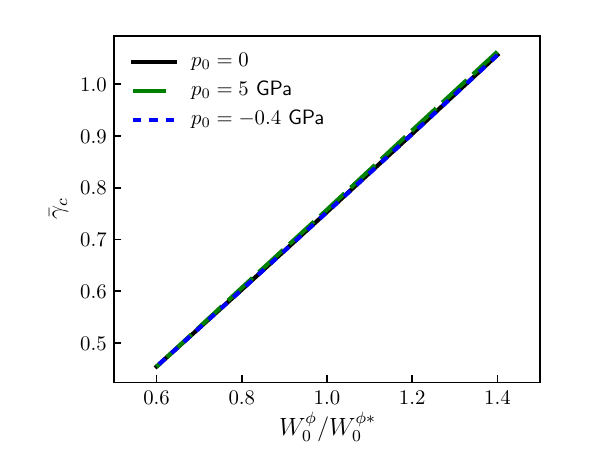} \label{fig6a}} \qquad
 \subfigure[defect vs.~threshold energy]{\includegraphics[width=0.32\textwidth]{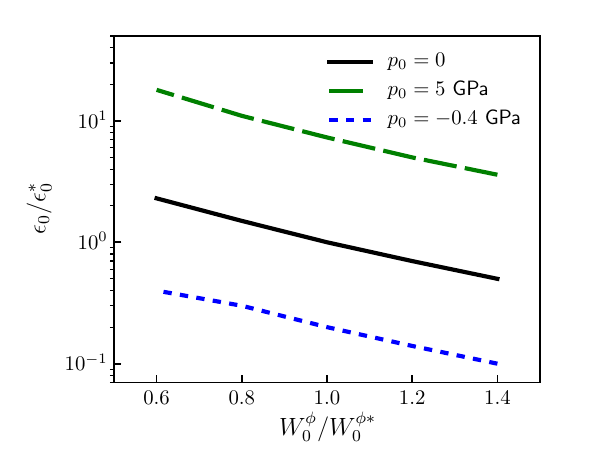}\label{fig6b}} \\
 \subfigure[defect vs.~cohesive energy]{\includegraphics[width=0.32\textwidth]{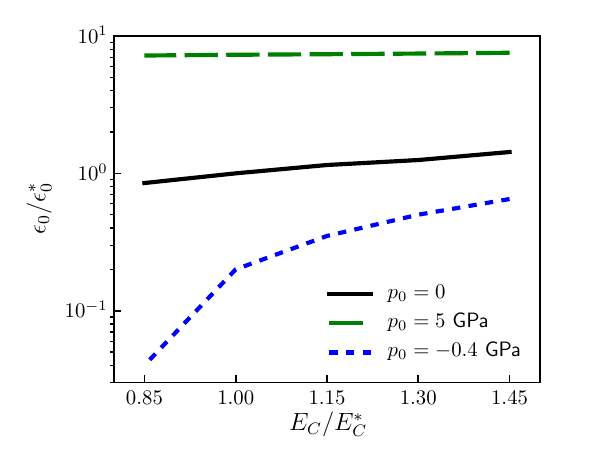}\label{fig6c}} \qquad
  \subfigure[average stress vs.~strain]{\includegraphics[width=0.32\textwidth]{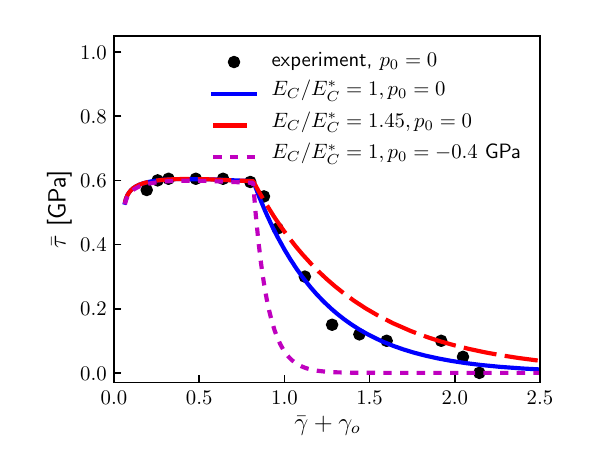}\label{fig6d}} 
  \end{center}  \vspace{-0.5cm}
\caption{Effects of threshold fracture energy $W^\phi_0$, cohesive energy $E_C$, and pressure $p_0$ on 
(a) average strain for localization $\bar{\gamma}_c$ and 
(b), (c) initial defect magnitude $\epsilon_0$ consistent with $\bar{\gamma_c}$
(d) average stress $\bar{\tau}$ vs.~strain $\bar{\gamma}$ at $W^{\phi}_0 = W^{\phi *}_0$
with experimental torsion data \cite{fellows2001b}.
Baseline ($W^{\phi *}_0, E_C^*, \epsilon^*_0$) from Table~\ref{table1}.}
\label{fig6}       
\end{figure}

For calculations reported in Fig.~\ref{fig6}, the full model with fracture and possible melting is invoked,
but melting never occurs in these particular predictions because fracture always takes place first.
As in \S5.2, upon shear fracture, stress drops too quickly to enable a large enough temperature rise  from stress power to initiate melting, even in the center of the shear band.

Effects of threshold energy $W^\phi_0$ on critical localization strain $\bar{\gamma}_c$,
at fixed $E_C$,
are shown in Fig.~\ref{fig6a}. Three values of applied pressure are considered: zero, 5 GPa compression,
and 0.4 GPa tension. Recall from \S5.1 that a tensile pressure approaching 0.5 GPa would induce
instant cavitation and rupture, so larger tensile pressures not modeled here.
The magnitude of each initial defect $\epsilon_0$ needed to achieve the corresponding
$\bar{\gamma}_c$ is shown in Fig.~\ref{fig6b}.
Critical average strain increases near linearly with $W^\phi_0$ but scantly increases (decreases) with
compressive (tensile) pressure at fixed $W^\phi_0$.
Initial defect intensity in Fig.~\ref{fig6b}, in contrast, decreases significantly as tensile stress increases
at fixed threshold $W^\phi_0$. Furthermore, perturbation strength must decrease to enable larger threshold energy
at fixed pressure. 
Results are physically viable: sensitivity to defects is greater in tension than compression,
and threshold energy drops for sharper defects.
Threshold energy and initial defects should not be prescribed independently because results show the two are intrinsically coupled.
The maximum local temperature in the core of the shear band, among all cases reported in Fig.~\ref{fig6}, arises for $W^\phi_0/W^{\phi *}_0 = 1.4$ and $p_0 = 5$ GPa, with a post-localization value
$\theta = 782$ K. Maximum core temperature increases
as $W^\phi_0$, $E_C$, and $p_0$ increase: threshold energy, cohesive energy,
and compressive stress all delay final failure from damage-softening and fracture.

Effects of cohesive energy on defect intensity needed to produce a consistent value of $\bar{\gamma}_c \approx 0.75$, at fixed $W^\phi_0 = W^{\phi *}_0$, are reported in Fig.~\ref{fig6c}.
At any value of $E_C$, a larger defect intensity $\epsilon_0$ is required in compression, a smaller defect strength in tension. Pressure sensitivity increases for lower values of $E_C$.
At fixed $p_0$, a larger $E_C$ requires a more severe defect to engender the same $\bar{\gamma}_c$, especially under tensile stress.
For $p_0 = 5$ GPa, $\bar{\gamma}_c = 0.757$, whereas $\bar{\gamma}_c = 0.753$
for $p_0 = 0$ and $-0.4$ GPa.
Cohesive energy and perturbation strength are nearly independent at high compressive pressure. 
Effects of tensile $p_0$ and increased $E_C$ are compared with experiment \cite{fellows2001b}
in Fig.~\ref{fig6d}, where $\epsilon_0$ is prescribed to give $\bar{\gamma}_c = 0.753$ in all cases shown.
Stress decay for $\bar{\gamma} > \bar{\gamma}_c$ 
accelerates under tensile pressure and decelerates when $E_C$ increases.

\begin{figure}
\begin{center}
 \subfigure[minimum localization strain]{\includegraphics[width=0.32\textwidth]{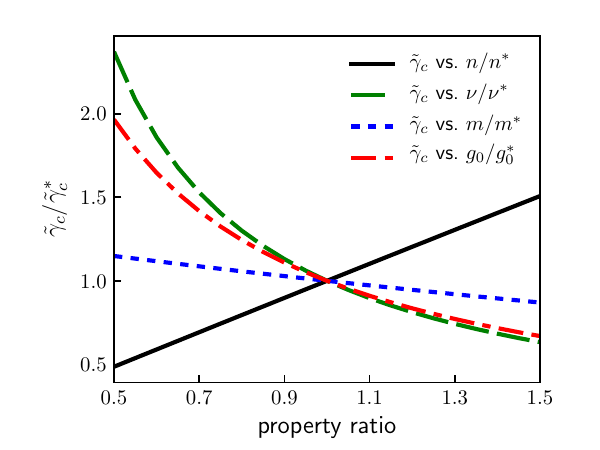} \label{fig7a}} \qquad
 \subfigure[minimum threshold energy]{\includegraphics[width=0.32\textwidth]{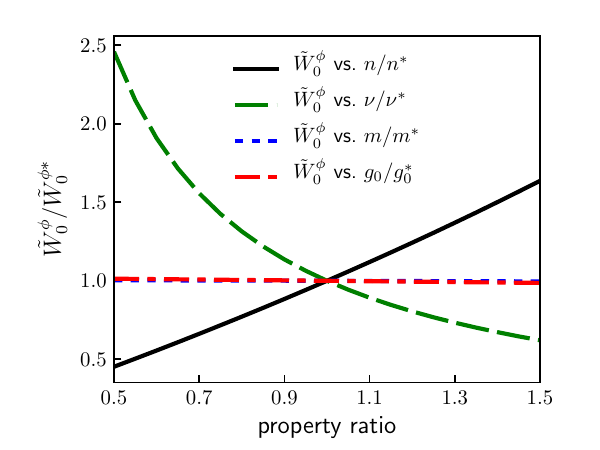}\label{fig7b}} 
  \end{center}
  \vspace{-0.5cm}
\caption{Effects of viscoplastic parameters for strain hardening $n > 0$, thermal softening $\nu < 0$,
rate sensitivity $m > 0$, and yield strength $g_0 > 0$ on
minimum localization (i.e., instability) strain $\tilde{\gamma}_c$ and corresponding
threshold energy $\tilde{W}^\phi_0$.
Baseline ($n^*,\nu^*,m^*,g^*_0$) from Table~\ref{table1}
produce $\tilde{\gamma}_c^* = 0.372$ and $\tilde{W}^{\phi *}_0 = 0.221$ GPa.}
\label{fig7}       
\end{figure}

Barring activation of some catastrophic defect or change in microstructure,
adiabatic shear localization should follow, but not precede, instability in a viscoplastic solid 
\cite{bai1982,anand1987,wright2002,fressengeas1987,molinari1988,shawki1988}.
Therefore, as the shear fracture model of \S3.4 seeks to represent material degradation
within a shear band, a pragmatic lower bound on the threshold plastic work $W^\phi_0$,
written as $\tilde{W}^\phi_0$, corresponds to the plastic work accumulated up to the
instability strain $\tilde{\gamma}_c$, that, in homogeneous simple shear, is
determined from the condition $ \d \tau / \d \gamma = 0$ under
adiabatic conditions at a given initial temperature $\theta_i$, loading rate
$\dot{\gamma} = \dot{\bar{\gamma}}$, and pressure $p_0$. 
This bound is inherent in prior modeling \cite{dolinski2010,dolinski2015b}
that used a similar threshold energy density for the onset of damage softening.
In the absence of melting, and with $\theta_i =  \theta_0$, the power-law flow
model of \eqref{eq:flowstress} with analytical temperature solution of \eqref{eq:thetanomelt} ($\gamma_f \rightarrow \gamma$)
gives the following implicit solution for minimum localization strain $\tilde{\gamma}_c$ and energy 
$\tilde{W}^\phi_c$:
\begin{align}
\label{eq:instrain}
&\tilde{\gamma}_{c} = \underset{\gamma \geq 0 }{\text{arg} \, 0} \left[ 
\frac{n}{\nu A_c}
+ \frac{(1 + \frac{ \gamma }{ \gamma_0})^{1+n}}{1 + A_c (\frac{1-\nu}{1+n}) \{ (1 + \frac{ \gamma }{ \gamma_0})^{1+n} - 1\} }
 \right], \qquad A_c = \frac{\beta_0 \chi_0 g_0 \gamma_0 }{c_V \theta_0} \biggr{(} \frac{\dot{\bar{\gamma}}}{ \dot{\gamma}_0} \biggr{)}^m;
 \\
 & \tilde{W}^\phi_0 = \int_0^{\tilde{\gamma}_c} \tau (\gamma, \theta(\gamma)) \, \d \gamma = \frac{c_V \theta_0}{\beta_0} \biggr{ [ } \biggr{ \{ }
 1 + A_c  \biggr{ ( } \frac{1-\nu}{1+n} \biggr{ ) } 
 \biggr{ [ } \biggr{ ( }  1 + \frac{\tilde{\gamma}_c}{ \gamma_0} \biggr{ ) }^{1+n} -1 \biggr{ ] }
 \biggr{ \} }^{\frac{1}{1-\nu} }
 -1 \biggr { ] }.
 \end{align}
 Parameters from Table~\ref{table1}, $\dot{\bar{\gamma}} = 3200$/s, and $p_0 = 0$ give
 $\tilde{\gamma}_c = 0.372$ and $\tilde{W}^\phi_0 = 0.221$ GPa.
 The former can be compared with Culver's critical strain $\hat{\gamma}_c = -\frac{n c_V \theta}{\nu \tau}$ (e.g., \cite{singh2021}),
 where $\hat{\gamma}_c = 0.306$ for calculated peak stress $\tau =  0.605$ GPa and $\theta = 351$ K.
 Effects of $n$, $\nu$, $m$, and $g_0$ on $\tilde{\gamma}_c$ and $\tilde{W}^\phi_0$ are shown in
 Fig.~\ref{fig7a} and Fig.~\ref{fig7b}.
 Both  $\tilde{\gamma}_c$ and $\tilde{W}^\phi_0$ increase nearly linearly with increasing strain hardening exponent $n$.
 Both decrease nonlinearly with decreasing (more strongly negative) $\nu$, corresponding to more thermal softening.
 While  $\tilde{\gamma}_c$ decreases modestly and nearly linearly with increasing
 rate sensitivity $m$, the minimum threshold energy $\tilde{W}^\phi_0$ is negligibly affected by $m$.
 The latter is likewise insensitive to initial static yield strength $g_0$, but $\tilde{\gamma}_c$ decreases
 nonlinearly with increasing $g_0$. The latter decrease is a result of proportionally more plastic dissipation and
 temperature rise,
 leading to thermal softening that overtakes strain hardening at a lower applied strain.
 
\begin{figure}[ht!]
\begin{center}
 \subfigure[critical strain vs.~melt]{\includegraphics[width=0.32\textwidth]{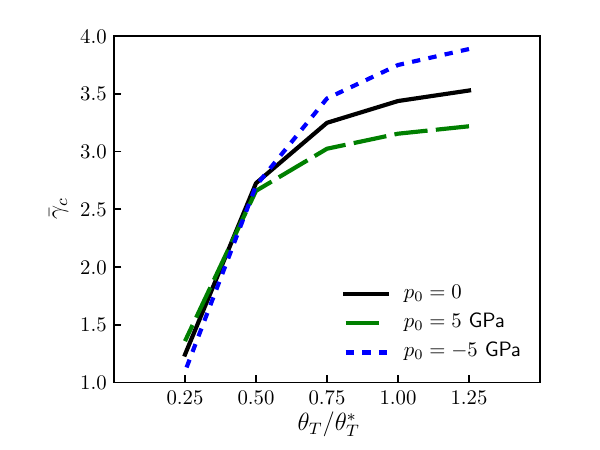} \label{fig8a}} 
 \subfigure[average stress vs.~strain]{\includegraphics[width=0.32\textwidth]{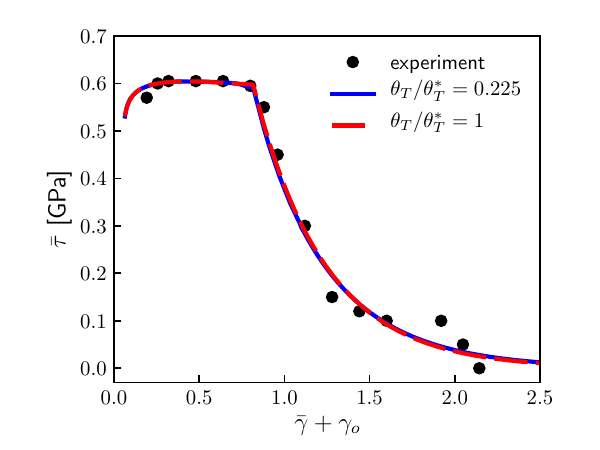}\label{fig8b}} 
 \subfigure[local shear strain]{\includegraphics[width=0.32\textwidth]{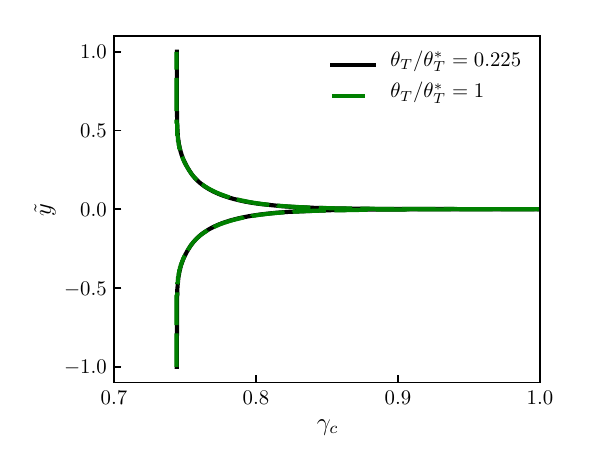}\label{fig8c}} \\
  \subfigure[local temperature]{\includegraphics[width=0.32\textwidth]{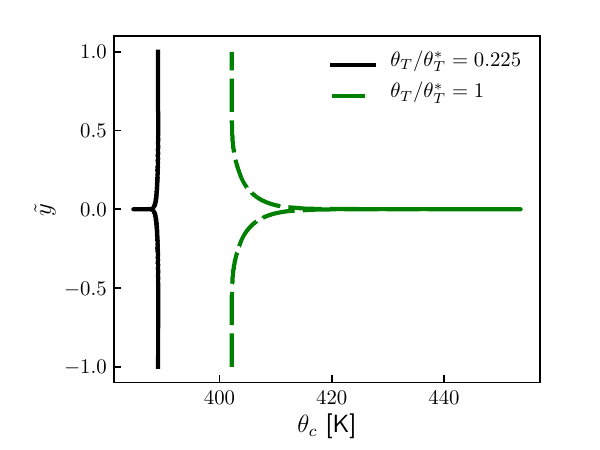} \label{fig8d}} 
 \subfigure[shear fracture]{\includegraphics[width=0.32\textwidth]{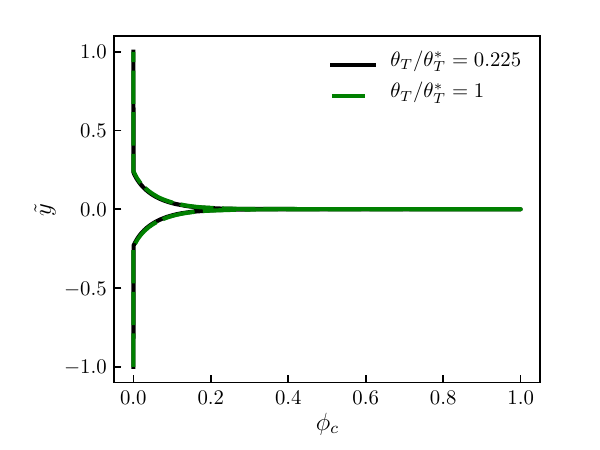}\label{fig8e}} 
 \subfigure[melt fraction]{\includegraphics[width=0.32\textwidth]{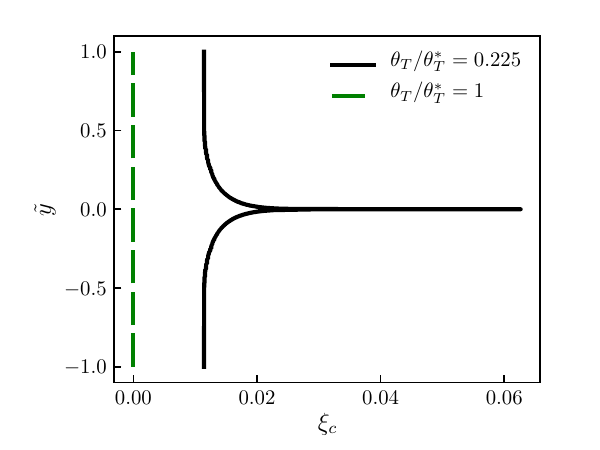}\label{fig8f}}
  \end{center}  \vspace{-0.5cm}
\caption{Model results for (a) effects of equilibrium melt temperature $\theta_T$ and pressure $p_0$
on critical localization strain $\bar{\gamma}_c$ with fracture suppressed (b)
average stress vs.~strain for reduced and realistic ($\theta_T^*$ from Table~\ref{table1}) melt temperatures (fracture enabled)
and experimental data \cite{fellows2001b} 
(c), (d), (e), (f) post-localization strain $\gamma_c$, temperature $\theta_c$, fracture parameter $\phi_c$, 
melt parameter $\xi_c$ for reduced and realistic melt temperatures}
\label{fig8}       
\end{figure}

Implications of varying the equilibrium melt temperature $\theta_T$ are also investigated.
Parametric calculations reported in Fig.~\ref{fig8a} study hypothetical cases when
fracture is suppressed (i.e., $W^\phi_0 \rightarrow \infty$). Here, the defect
profile is fixed at nominal values from Table 1: $\epsilon_0 = 10^{-3}, \lambda_0 = 0.5$.
Influences of compressive or tensile pressure $p_0 = \pm 5$ GPa are considered,
where large tensile pressure is admissible since fracture and cavitation are suppressed.
Melt temperature $\theta_T$ ranges from 450
to 2250 K in Fig.~\ref{fig8a}.
In all cases, $\bar{\gamma}_c$ increases with increasing $\theta_T$ at fixed $p_0$:
loss of shear strength is delayed as the requisite temperature rise for melting increases.
In Fig.~\ref{fig8a}, $\bar{\gamma}_c$ decreases (increases) in tension (compression) at low $\theta_T$, with the opposite trend at high $\theta_T$. The effect of $p_0$ is greater at high $\theta_T$.
Opposing effects ensue: compression resists melting due to $\bar{p}_0 \delta^\xi_0 > 0$ in \eqref{eq:xisol2}, but $p_0$ 
increases strength via \eqref{eq:gyp} and thus dissipative temperature rise that promotes melting.

Lastly, a theoretical case in which fracture and melting take place concurrently is studied.
In this case, $\theta_T = 0.225 \theta_T^* = 405$ K is low enough that melting begins in the core of the shear
band even when the fracture model is enabled.
Here, all parameters except $\theta_T$ and $\epsilon_0$ are nominal values from Table~\ref{table1}, including the set
($\lambda_0 = 0.5, \beta_\xi = 1.21\,$GPa$, r_0^\xi = 1$).
The value $\epsilon_0 = 7 \times 10^{-4}$ is reduced from $10^{-3}$ so
that localization always ensues at $\bar{\gamma}_c = 0.753$. 
Average stress-strain behavior differs little in Fig.~\ref{fig8b}
between the nominal case and reduced melting temperature.
Contours of local shear strain $\gamma_c$ and damage order parameter $\phi_c$ are also hardly affected
by $\theta_T$ in Figs.~\ref{fig8c} and \ref{fig8e}.
The material only partially melts, with nonzero $\xi_c$ limited to a narrow region near the core of the
shear band in Fig.~\ref{fig8f}.  Temperature $\theta_c$ in Fig.~\ref{fig8d} decreases slightly, relative to surrounding material, in the core
where this transition occurs, as thermal energy is consumed by latent heat.

\section{Conclusions \label{sec7}}

An analytical-numerical framework has been advanced to study
interplay among adiabatic shear localization, ductile fracture, and melting
in viscoplastic solids.
The problem studied is adiabatic simple shear with uniform external pressure.
Localization initiates when
 a threshold energy density from cumulative plastic work is attained
 in the context of a dynamic shear fracture model.
 Post-localization stress decay is
 modulated by the cohesive energy (related to the shear-band fracture
 toughness), with the decay rate increased (decreased) by tensile (compressive) stress.
 Localization concludes when a band of infinitesimal thickness and infinite shear
strain is admitted by solutions to the governing equations.
Threshold energy, initial defects (e.g., local strength perturbations), and critical average 
strain at which localization begins are not independent.
Larger threshold energy implies a larger critical strain and a defect of
lesser intensity or broader distribution.
Sharp intense defects are most deleterious; 
defect sensitivity increases with tensile pressure.
Following the usual convention that instability is a necessary condition for localization, the threshold 
energy logically increases with strain hardening propensity and decreases with more
thermal softening.

The model is exercised using properties for a high-strength
steel (i.e., RHA). Calculated results for average stress-strain behavior, including post-localization
response, accurately match experimental data from dynamic torsion using
a literature value of threshold energy density for steel.  Nearly all parameters are obtained
from cited research. Calibration is required
only for initial yield strength and cohesive energy. The latter's value is within bounds suggested
by experiments and prior phase-field studies. 
Model features and parameters, with fracture enabled, limit accurate calculations to maximum compressive pressures of 5 GPa and tensile pressures under 0.5 GPa.

The framework idealizes the fully formed core of a shear band as a singular region of
infinitesimal width and infinite strain. Thus, experimentally measured widths and 
magnitudes of the band in a continuous sample are necessarily under- and over-predicted, respectively,
when experimental profiles do not account for singularities (e.g., shear cracks).
Realistic absolute widths cannot be calculated since the analysis
includes no material length scale: presently calculated widths are proportional to specimen size for an a priori, normalized initial perturbation profile.
To accurately calculate the absolute width, a complete theoretical model with
heat conduction, inertia, and phase-field (i.e., damage order-parameter) gradient regularization 
should be implemented in a numerical scheme with 
full space-time discretization.
The governing equations for such a complete theoretical model have already
been included in the theory outlined in \S2 of this work, but certain material
constants have been zeroed out to allow use of the infinite-strain, singular-point
localization criterion pioneered by Molinari and Clifton. 
High defect sensitivity inherent in this criterion should be reduced in a regularized theoretical-numerical setting:
 a stronger initial perturbation is expected to be needed to overcome energetic resistance
 from presently omitted regularization mechanisms.
 
 Calculations suggest that melting is unimportant for shear localization this steel under dynamic
 torsion at rates on the order of $10^4$/s. Shear fracture progresses faster than the temperature rise needed for melt initiation.
 Maximum predicted temperatures in the core of the band are presumably elevated due to omission of thermal diffusion; however,
 it is possible that the intentionally simple damage model implemented herein omits dissipative structural processes that could competitively increase local temperature.
 The temperature rise within the shear-softened zone likewise appears insufficient
 for $\alpha \rightarrow \gamma$ phase transition in this steel.
 It is not impossible than an $\alpha \rightarrow \epsilon$ transformation
 could occur, but this transformation has not been confirmed in known experiments
 on RHA at low pressures (e.g., $p \leq 5$ GPa as studied here).
 If much higher pressures, or if metals with greater intrinsic resistance to local shear fracture
 (i.e., much larger threshold or cohesive energies), are considered, then higher temperatures can be attained. Therefore, the present inferences of low importance of melting and phase transitions should not be blindly extrapolated to more extreme pressure conditions, other (e.g., multi-axial) stress-strain histories, or to all kinds of steels.

\footnotesize
\bibliography{refs}
%
%
\end{document}